\documentclass[apj]{emulateapj}

\newcommand\bibinc{n}		

\usepackage{subeqnarray}
\usepackage{natbib}
\usepackage{color}
\usepackage[utf8]{inputenc}

\DeclareMathSymbol{\varOmega}{\mathord}{letters}{"0A}
\DeclareMathSymbol{\varSigma}{\mathord}{letters}{"06}
\DeclareMathSymbol{\varPsi}{\mathord}{letters}{"09}

\newcommand{\App}[1]{Appendix~\ref{#1}}

\definecolor{gray}{gray}{0.5}

\begin{document}


\title{C/O and Snowline Locations in Protoplanetary Disks: The Effect of Radial Drift and Viscous Gas Accretion}

\author{Ana-Maria A. Piso\altaffilmark{1}, Karin I. \"Oberg\altaffilmark{1}, Tilman Birnstiel\altaffilmark{1, 3}, Ruth A. Murray-Clay\altaffilmark{2}}
\altaffiltext{1}{Harvard-Smithsonian Center for Astrophysics, 60 Garden Street, Cambridge, MA 02138}
\altaffiltext{2}{Department of Physics, University of California, Santa Barbara, CA 93106}
\altaffiltext{3}{Max Planck Institute for Astronomy, K\"onigstuhl 17, 69117 Heidelberg, Germany}

\begin{abstract}
The C/O ratio is a defining feature of both gas giant atmospheric and protoplanetary disk chemistry. In disks, the C/O ratio is regulated by the presence of snowlines of major volatiles at different distances from the central star. We explore the effect of radial drift of solids and viscous gas accretion onto the central star on the snowline locations of the main C and O carriers in a protoplanetary disk, H$_2$O, CO$_2$ and CO, and their consequences for the C/O ratio in gas and dust throughout the disk. We determine the snowline locations for a range of fixed initial particle sizes and disk types. For our fiducial disk model, we find that grains with sizes $\sim$$0.5$ cm $\lesssim s \lesssim$ 7 m for an irradiated disk, and $\sim$$0.001$ cm $\lesssim s \lesssim$ 7 m for an evolving and viscous disk, desorb at a size-dependent location in the disk, which is independent of the particle's initial position. The snowline radius decreases for larger particles, up to sizes of $\sim$7~m. Compared to a static disk, we find that radial drift and gas accretion in a viscous disk move the H$_2$O snowline inwards by up to 40 \%, the CO$_2$ snowline by up to 60 \%, and the CO snowline by up to 50 \%. We thus determine an inner limit on the snowline locations when radial drift and gas accretion are accounted for. 
\end{abstract}

\section{Introduction}
\label{sec:intro}



The chemical composition of protoplanetary disks affects planet formation efficiencies and the composition of nascent planets.
Gas giants accrete their envelopes from the nebular gas. As such, planet compositions are tightly linked to the structure and evolution of the protoplanetary disk in which they form. It is thus essential to understand the disk chemistry and dynamics well enough to (1) predict the types of planet compositions that result from planet formation in different parts of the disk, and (2) backtrack the planet formation location based on planet compositions. 




The structures of protoplanetary disks are complex, and affected by a multitude of chemical and dynamical processes (see review by \citealt{henning13}). From the chemistry perspective, volatile compounds are particularly important. Their snowline locations determine their relative abundance in gaseous and solid form in the disk,. Based on protostellar and comet abundances, some of the most important volatile molecules are H$_2$O, CO$_2$, CO, N$_2$. Recent observations of protoplanetary disks have provided valuable information about the abundances and snowline locations of some of these compounds. For example, the CO snowline has been detected in the disk around TW Hya \citep{qi13}, as well as in the disk around HD 163296 \citep{mathews13} using line emissions from DCO$^+$. Observations of TW Hya have also revealed a H$_2$O snowline \citep{zhang13}, and more such snowline detections are expected in future ALMA cycles. These observations are currently lacking an interpretive framework that takes into account all important dynamical and chemical processes. Furthermore, such a framework is crucial to connect observed snowline locations to planet formation.



An important consequence of snowline formations in disks is that disks are expected to present different carbon-to-oxygen (C/O) ratios in the gas and in icy dust mantles at different disk radii. This effect was quantified by  \citet{oberg11}, who considered the fact that the main carries of carbon and oxygen, i.e. H$_2$O, CO$_2$ and CO, have different condensation temperatures. This changes the relative abundance of C and O in gaseous and solid form as a function of the snowline location of the volatiles mentioned above. \citet{oberg11} calculated analytically the C/O ratio in gas in dust as a function of semimajor axis for passive protoplanetary disks and found a gas C/O ratio of order unity between the CO$_2$ and CO snowlines, where oxygen gas is highly depleted. This effect was used to explain claims of detections of superstellar C/O ratios in exoplanet atmospheres (e.g., WASP-12b, \citealt{madhu11}), which however have been unambiguously refuted (\citealt{stevenson14}, \citealt{kreidberg15}).

\citet{oberg11} assumed a static disk with no chemical evolution. In reality, dynamical and chemical processes affect the snowline locations and the resulting C/O ratio. Several works have addressed some of these effects. \citet{madhu14} use a steady-state active disk model that includes planetary migration and use the C/O ratio to constrain migration mechanisms. \citet{alidib14} calculate the C/O ratio throughout the disk by incorporating the evolution of solids, i.e. radial drift, sublimation and grain coagulation, as well as the diffusion of volatile vapors. \citet{alidib14} use the 1+1D $\alpha$-disk model of \citet{hughes10}, in which the gas drifts outwards in the disk midplane, and thus small particles that are well-coupled to the gas will also advect outward. Their model assumes a cyclical conversion between H$_2$O or CO dust and vapor: large enough particles that are decoupled from the gas drift inwards and start desorbing. Once their sublimation is complete, back-diffusion moves the H$_2$O or CO vapor ouwards to their respective snowlines, where they instantly condense into mm-sized particles that diffuse outwards with the gas while coagulating into larger particles. Once the grains become large enough to decouple from the gas and drift inwards, the cycle restarts. This ```conveyor belt''' model is based on the pioneering work by \citet{cuzzi04}, and \citet{ciesla06} for the evolution of H$_2$O in a viscous disk. This approach leads \citet{alidib14} to find that the gaseous C/O ratio increases with time inside the H$_2$O snowline, approaching unity at 2 AU after $\sim$$10^4-10^5$ years.  \citet{thiabaud15} consider additional carbon carrier volatile species in their chemical network, such as CH$_4$, and find that the gas C/O ratio may be enriched by up to four times the Solar value in the outer parts of the disk where CH$_4$ and CO are the only gaseous carriers of C and O.  They also include nitrogen carriers such as N$_2$ or NH$_3$, and perform similar calculations for nitrogen. 

Each of these studies have considered a specific combination of dynamical and chemical effects. One scenario that has not yet been considered is the combination of radial drift and viscous gas accretion in isolation. Studying these two dynamical processes makes it possible to quantify their separate effect on snowline locations and the C/O ratio at various disk radii.

In this paper, we perform a systematic study to understand the detailed qualitative and quantitative effects of radial drift and gas accretion on the H$_2$O, CO$_2$ and CO snowline locations, and the resulting C/O ratio in gas and dust throughout the protoplanetary disk. More importantly, we obtain a limit on how close to the star the snowline locations can be pushed by radial drift and gas accretion.

This paper is organized as follows. In Section \ref{sec:model}, we present our disk, radial drift and desorption models, as well as the timescales relevant to the coupled drift-desorption process. We calculate the H$_2$O, CO$_2$ and CO snowline locations as a function of particle size for an irradiated and an evolving disk in Section \ref{sec:snowlines}, and the resulting C/O ratio throughout the disk in Section \ref{sec:COratio}. In Section \ref{sec:discussion}, we discuss the generality of our results, as well as additional effects on the snowline locations. Finally, we summarize our findings in Section \ref{sec:summary}.

\section{Model Framework} 
\label{sec:model}

We present our protoplanetary disk model for a static, an irradiated, an evolving, and a viscous disk in section \ref{sec:disk}. In section \ref{sec:drift}, we describe our analytic model for the radial drift of solids. We summarize our ice desorption model in section \ref{sec:desorption}. Finally, we discuss the relevant timescales for dynamical effects that affect snowline locations 
in section \ref{sec:timescales}.

\subsection{Disk Model}
\label{sec:disk}

To understand the separate effects of radial drift, radial movement of gas throughout the disk due to gas accretion, and accretion heating, we use four separate disk models:  \textit{static disk}, which is solely irradiated by the host star and does not take into account gas accretion onto the star or radial drift; \textit{irradiated disk}, which has the same temperature profile as the static disk and does not experience gas accretion or accretional heating, but it takes into account radial drift of solids; \textit{evolving disk}, in which the gas is accreting onto the central star causing the gas surface density to decrease with time, but which does not experience accretion heating; and \textit{viscous disk}, for which the mass flux $\dot{M}$ is constant in time and independent of semimajor axis, and the temperature profile is calculated using both accretional heating and stellar irradiation.


\textbf{Static and Irradiated disk.} We adopt a minimum mass solar nebula (MMSN) disk model for a static and and an irradiated disk similar to the prescription of \citet{chiang10}. The gas surface density and midplane temperature are
\begin{subeqnarray}
\label{eq:disk}
\Sigma&=&2000\, (r/\text{AU})^{-1}\,\, \text{g cm}^{-2} \slabel{eq:disksigma}\\
T &=& 120\, (r/\text{AU})^{-3/7} \,\,\text{K}, \slabel{eq:diskT}
\end{subeqnarray}
where $r$ is the semimajor axis. Our surface density profile is flatter than the $\Sigma \propto r^{-3/2}$ used by \citet{chiang10}. Our choice is inspired by observations of protoplanetary disks at radii larger than $\sim$20 AU (e.g., \citealt{andrews10}), which suggest that typical disks may have surface density profiles with $\Sigma \propto r^{-1}$. A slope flatter than $\Sigma \propto r^{-3/2}$ is also more consistent with the temperature profile for a steady-state gas disk (see the Viscous disk heading below and \App{app:steadystate}). We use the static disk model to compare our results with those of \citet{oberg11}.


\textbf{Evolving disk.} We model the evolving disk as a thin disk with an $\alpha$-viscosity prescription \citep{shakura73}:
\begin{equation}
\label{eq:nu}
\nu=\alpha c H.
\end{equation}
Here $\nu$ is the kinematic viscosity, $\alpha < 1$ is a dimensionless coefficient and we choose $\alpha=0.01$, and $c$, $H$ are the isothermal sound speed and disk scale height, respectively:
\begin{subeqnarray}
\label{eq:cdHd}
c &=& \sqrt{\frac{k_{\rm B} T}{\mu m_{\rm p}}} \slabel{eq:cd} \\
H&=& \frac{c}{\Omega_{\rm k}} \slabel{eq:Hd},
\end{subeqnarray}
where $k_{\rm B}$ is the Boltzmann constant, $\mu$ is the mean molecular weight of the gas, $m_{\rm p}$ is the proton mass, and $\Omega_{\rm k} \equiv \sqrt{G M_*/r^3}$ is the Keplerian angular velocity,  with $G$ the gravitational constant and $M_*$ the stellar mass. We choose $M_*=M_{\odot}$ and $\mu=2.35$, corresponding to the Solar composition of hydrogen and helium. The temperature profile for the 
evolving disk is assumed to be the same as for the irradiated disk and given by Equation (\ref{eq:diskT}). From Equations (\ref{eq:nu}) and (\ref{eq:cdHd}), the viscosity can thus be expressed as a power-law in radius, $\nu \propto r^{\gamma}$, with $\gamma=15/14 \approx 1$ for our choice of parameters. Following \citet{hartmann98}, we define $R \equiv r/r_{\rm c}$ and $\nu_{\rm c} \equiv \nu(r_{\rm c})$, where $r_{\rm c}$ is a characteristic disk radius. We choose $r_{\rm c}=100$ AU. The gas surface density is given by the self-similar solution

\begin{equation}
\label{eq:Sigmaact}
\Sigma(R, \tilde t) = \frac{M (2 - \gamma)}{2 \pi r_{\rm c}^2 R^{\gamma}} \tilde t^{-(5/2-\gamma)/(2-\gamma)} \exp{\Big[-\frac{R^{(2-\gamma)}}{\tilde t}}\Big],
\end{equation}
where $M$ is the total disk mass and
\begin{subeqnarray}
\label{eq:T}
\tilde t & \equiv & \frac{t}{t_{\rm c}} + 1 \\
t_{\rm c} & \equiv & \frac{1}{3(2-\gamma)^2} \frac{r_{\rm c}^2}{\nu_{\rm c}},
\end{subeqnarray}
where $t$ is time. We choose $M=0.1 M_{\odot}$ (e.g., \citealt{birnstiel12}), but we note that our results are insensitive to this choice (see Section \ref{sec:discussion}).  The irradiated and evolving disk surface densities match at $t \approx 5 \times 10^5$ years in the inner disk, but they diverge at distances larger than a few AU due the exponential cutoff in radius of the surface density of the evolving disk (Equation \ref{eq:Sigmaact}).

\textbf{Viscous disk.} 
Calculating the midplane temperature self-consistently for an evolving disk that is also actively heated, and thus whose thermal evolution is dominated both by accretion heating and stellar irradiation, is non-trivial. We therefore use instead the Shakura-Sunyaev thin disk steady-state solution to derive the midplane temperature profile, $T_{\rm act}$. The equations governing the evolution of the steady-state disk are listed in \App{app:steadystate}. We assume an interstellar opacity for the dust grains given by \citet{bell94}, but reduced by a factor of 100. This reduction is due to the fact that disk opacities are lower than the interstellar one. While this scaling is consistent with more detailed models of grain opacities in disks (e.g., \citealt{mordasini14}), realistic disk opacities are much less sensitive to changes in temperature than the interstellar opacity if substantial grain growth has occurred. However, the disk temperature does not vary significantly across the small region of the disk where accretion heating is important ($r \lesssim 1$ AU). Moreover, using an analytic opacity formula is more convenient since it results in a constant gas surface density in the inner disk region (see below). Our opacity law is thus

\begin{equation}
\label{eq:opacity}
\kappa=\kappa_0 T_{\rm act}^2,
\end{equation}
where $\kappa_0=2 \times 10^{-6}$. By solving the Equation set (\ref{eq:diskeq}) we find
\begin{equation}
\label{eq:Tdact}
T_{\rm act}=\frac{1}{4 r} \Big(\frac{3 G \kappa_0\dot{M}^2 M_* \mu m_{\rm p} \Omega_{\rm k}}{\pi^2 \alpha k_{\rm B} \sigma}\Big)^{1/3}.
\end{equation}
Since both accretion heating and stellar irradiation contribute to the thermal evolution of the disk, we compute the midplane temperature for our viscous disk as
\begin{equation}
\label{eq:activeT}
T^4 = T_{\rm act}^4 + T_{\rm irr}^4,
\end{equation}
where to avoid notation confusion $T_{\rm irr}=T$ from Equation (\ref{eq:diskT}), the temperature profile for an irradiated disk. We can then easily determine  $c$ and $H$ from Equation (\ref{eq:cdHd}), as well as the viscosity $\nu$ from Equation (\ref{eq:nu}) for a given $\alpha$. For consistency, we choose  $\alpha=0.01$ as in the previous case. Finally, we determine $\Sigma$ from Equation (\ref{eq:Mdot}), where we choose $\dot{M}=10^{-8} M_{\odot}$ yr$^{-1}$ based on disk observations (e.g., \citealt{andrews10}). In the inner portion of our disk ($r  \lesssim 1$ AU for our fiducial model with $\dot{M}=10^{-8} M_{\odot}$ yr$^{-1}$), our choice of opacity (Equation \ref{eq:opacity}) implies that the disk has a constant surface density with radius (see Equations \ref{eq:diskeq}).

Before we proceed forward, we note that our disk models assume a constant stellar luminosity $L_*$, as well as a constant mass accretion rate $\dot{M}$ for the viscous disk. In reality, the stellar luminosity decreases as the host star contracts, which will reduce the disk temperature and push the snowlines inward, as we explain in Section \ref{sec:neglected}. For a Solar type star, as our fiducial model assumes, $L_*$ remains relatively constant during the star's pre-main sequence evolution of $\sim$ 10 Myr \citep{kennedy06}, which is larger than the giant planet formation timescale. Thus the midplane temperature will not change significantly for our model due to variations in stellar luminosity, but it may decrease substantially for smaller stars, pushing the snowline inward (see Section \ref{sec:neglected} for details). Realistic mass accretion rates, $\dot{M}$, may vary between $\sim$$10^{-7}$ and $\sim$$10^{-9}$ $M_{\odot}$ yr$^{-1}$ as the disk evolves (e.g., \citealt{chambers09}, \citealt{sicilia10}). For $\dot{M}\lesssim 10^{-9} M_{\odot}$ yr$^{-1}$, the disk becomes optically thin and hence depleted of gas, which means giant planets must have formed before $\dot{M}$ becomes too low. \citet{garaud07} find that the snowline locations scale as $r_{\rm snow} \propto \dot{M}^{1/3}$. A factor of 100 reduction in the mass accretion rate will thus move the H$_2$O snowline inwards by a factor of $\sim$$4$ --- since accretion heating is dominant only in the inner disk, the CO$_2$ and CO snowline locations are unlikely to be affected by changes in mass accretion rate. The inward movement of the H$_2$O snowline due to the decrease in $\dot{M}$ may be even larger, by up to one order of magnitude \citep{chambers09}. We thus conclude that changes in $L_*$ throughout time may only modestly affect our results, while changes in $\dot{M}$ may significantly affect our results for the H$_2$O snowline, as its location may be determined by the decline in mass accretion rate rather than radial drift. The time variability of $L_*$ and $\dot{M}$ should be taken into account when drawing more robust conclusions, as well as for different host star and disk properties.

\subsection{Radial Drift}
\label{sec:drift}

Solid particles orbit their host star at the Keplerian velocity $v_{\rm k} \equiv \Omega_{\rm k} r$. The gas, however, experiences an additional pressure gradient, which causes it to rotate at sub-Keplerian velocity \citep{weidenschilling77}. Dust grains that are large enough thus experience a headwind, which removes angular momentum, causing the solids to spiral inwards and fall onto the host star. Small particles are well-coupled to the gas, while large planetesimals are decoupled from the gas. From the review by \citet{chiang10}, the extent of coupling is quantified by the dimensionless stopping time, $\tau_{\rm s} \equiv \Omega_{\rm k} t_{\rm s}$, where $t_{\rm s}$ is
\begin{equation}
\label{eq:ts}
t_{\rm s}= \left\{
\begin{array}{l l}
\rho_{\rm s} s / (\rho c), & \quad s < 9 \lambda/4 \,\,\,\ \text{Epstein drag} \\
4 \rho_{\rm s} s^2 / (9 \rho c \lambda), & \quad s < 9 \lambda/4, \,\text{Re} \lesssim 1 \,\,\,\ \text{Stokes drag.}
\end{array} 
\right.
\end{equation}
Here $\rho$ is the gas midplane density, $\rho_{\rm s}=2$ g cm$^{-3}$ is the density of a solid particle, $s$ is the particle size, $\lambda$ is the mean free path, and Re is the Reynolds number. 

For an irradiated disk, the radial drift velocity can be approximated as
\begin{equation}
\label{eq:rdotpas}
\dot{r} \approx -2 \eta \Omega_{\rm k} r \Big(\frac{\tau_{\rm s}}{1+\tau_{\rm s}^2}\Big),
\end{equation}
where
\begin{equation}
\label{eq:eta}
\eta \equiv - \frac{\partial P/\partial \ln r}{2 \rho v_{\rm k}^2} \approx \frac{c^2}{2 v_k^2}
\end{equation}
and $P = \rho c^2$ is the disk midplane pressure. 

For an evolving disk, the radial drift velocity has an additional term due to the radial movement of the gas \citep{birnstiel12}, i.e.
\begin{equation}
\label{eq:rdotact}
\dot{r} \approx -2 \eta \Omega_{\rm k} r \Big(\frac{\tau_{\rm s}}{1+\tau_{\rm s}^2}\Big) + \frac{\dot{r}_{\rm gas}}{1+\tau_{\rm s}^2},
\end{equation}
where $\dot{r}_{\rm gas}$ is the radial gas accretion velocity and can be expressed as (e.g., \citealt{fkr02})
\begin{equation}
\label{eq:vgas}
\dot{r}_{\rm gas} = - \frac{3}{\Sigma \sqrt{r}} \frac{\partial}{\partial r}(\nu \Sigma \sqrt{r}) 
\end{equation}
with $\Sigma$ from Equation (\ref{eq:Sigmaact}). For the viscous disk (see Section \ref{sec:disk}), $\dot{r}_{\rm gas}$ can be expressed more simply using the definition of the mass flux, $\dot{M}=-2 \pi r \dot{r}_{\rm gas} \Sigma$, with $\dot{M}$ fixed and $\Sigma$ obtained from Equation (\ref{eq:Mdot}). For our choice of parameters for both the evolving and the viscous disks, we have found that the radial flow of gas(calculated from Equation \ref{eq:vgas} for the evolving disk and from $\dot{r}_{\rm gas}=-\dot{M}/(2 \pi r \Sigma)$ for the viscous disk) is always directed inward for our parameter space of interest, in contrast with the model of \citet{alidib14} which assumes that the gas drifts outwards (see Section \ref{sec:intro}). For our evolving disk, the gas starts drifting outwards at a radius $r_{\rm switch} \approx 200 AU$, which is however well outside the CO snowline in our model. We thus note that variations in our fiducial disk model parameters (e.g., $T$, $\Sigma$, $\dot{M}$) may cause the gas to flow outwards in the outer parts of the disk, specifically at the CO$_2$ and CO snowline locations. Since drifting particles larger than a few cm are only modestly affected by gas accretion, an outward gas flow would move the CO$_2$ and CO locations further away from the star only for the smallest particles in our model, which are well-coupled to the gas.

\subsection{Volatile Desorption}
\label{sec:desorption}


In order for a volatile species to thermally desorb, it has to overcome the binding energy that keeps it on the grain surface. Following \citet{hollenbach09}, the desorption rate per molecule for a species $x$ can be expressed as
\begin{equation}
\label{eq:Rdes}
R_{\rm{des}, x} = \nu_x \exp{(-E_x/T_{\rm grain})},
\end{equation}
where $E_x$ is the adsorption binding energy in units of Kelvin, $T_{\rm grain}$ is the grain temperature, and $\nu_x=1.6 \times 10^{11} \sqrt{(E_x/\mu_x)}$ s$^{-1}$ is the molecule's vibrational frequency in the surface potential well, with $\mu_x$ the dimensionless mean molecular weight. We assume that the dust and gas have the same temperature in the disk midplane, hence $T_{\rm grain}=T$. For H$_2$O, CO$_2$ and CO, the binding energies $E_x$ are assumed to be 5800 K, 2000 K and 850 K, respectively (\citealt{collings04}, \citealt{fraser01}, \citealt{aikawa96}). We use the desorption rate, $R_{\rm des}$, to estimate the desorption timescale for particles of different sizes as described in section \ref{sec:timescales}.


\subsection{Relevant Timescales}
\label{sec:timescales}

We can estimate the extent to which radial drift and gas accretion affect desorption by comparing the timescales for desorption, drift and accretion, for solids of different sizes and compositions. 

\textit{Desorption timescale.} We assume that the solid bodies are perfect spheres and are entirely composed of only one volatile species, i.e. either H$_2$O, CO$_2$ or CO \footnote{We discuss the validity of these simplifications in section \ref{sec:discussion}.}. The timescale to desorb a single layer of molecules can then be estimated as
\begin{equation}
\label{eq:tdes}
t_{\rm des}=\frac{\rho_{\rm s}}{3 \mu_x m_{\rm p}} \frac{s}{N_x R_{\rm des, x}},
\end{equation}
where $N_x \approx 10^{15}$ sites cm$^{-2}$ is the number of adsorption sites of volatile $x$ per cm$^2$, assuming that the particle has a smooth surface \citep{hollenbach09}. 

\textit{Radial drift timescale.} To order of magnitude, the radial drift timescale can be estimated as 
\begin{equation}
\label{eq:tdrift}
t_{\rm drift} \sim \Big|\frac{r}{\dot{r}}\Big|,
\end{equation}
where $\dot{r}$ is the radial drift velocity given by Equation (\ref{eq:rdotpas}) for an irradiated disk and by Equation (\ref{eq:rdotact}) for an evolving disk.

\textit{Gas accretion timescale.} The timescale for gas accretion onto the central star for an evolving disk is (e.g., \citealt{armitage10})
\begin{equation}
\label{eq:tgas}
t_{\rm gas, acc} \sim \frac{r^2}{\nu} \sim \frac{1}{2 \alpha \eta \Omega_{\rm k}},
\end{equation}
with the latter expression derived from Equations (\ref{eq:nu}) and (\ref{eq:eta}).

\begin{figure}[h!]
\centering
\includegraphics[width=0.5\textwidth]{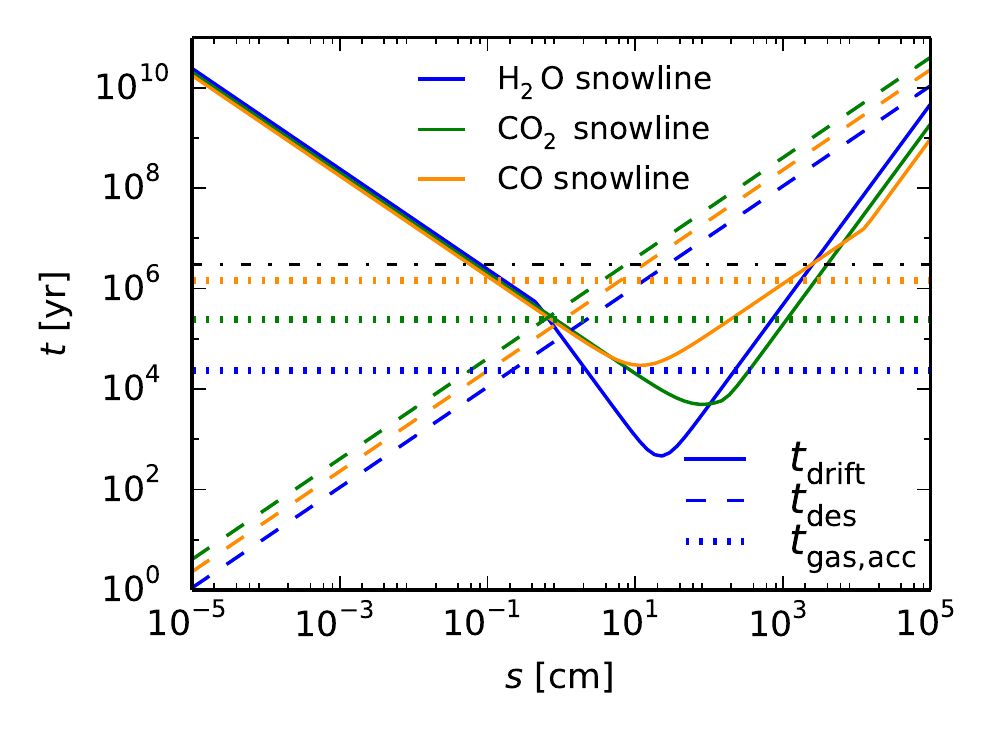}
\caption{Relevant timescales for dynamical effects in the desorption process: $t_{\rm drift}$ (solid lines), $t_{\rm des}$ (dashed lines) and $t_{\rm gas, acc}$ (dotted lines). The timescales are calculated at three representative locations, i.e. the H$_2$O, CO$_2$ and CO snowlines in the static disk. For our choice of parameters, the snowlines are located at $\sim$0.7 AU (blue lines), $\sim$8.6 AU (green lines) and $\sim$59 AU (red lines), respectively. The horizontal dot-dashed line represents a typical disk lifetime of 3 Myr. The particle size ordering at the minimum $t_{\rm drift}$ is not monotonic in snowline distance due to different drag regimes for those particle sizes at the snowline locations (Epstein drag at the H$_2$O and CO$_2$ snowlines, and Stokes drag at the CO snowline). Similarly, the ordering of $t_{\rm des}$ is not monotonic in snowline distance due to the non-monotony in mean molecular weight between H$_2$O, CO$_2$ and CO (18 $m_{\rm p}$, 44 $m_{\rm p}$ and 28 $m_{\rm p}$, respectively). Radial drift and gas accretion affect desorption in the regions where their respective timescales, i.e. $t_{\rm drift}$ and $t_{\rm gas, acc}$, are comparable to the desorption timescale $t_{\rm des}$.} 
\label{fig:timescales}
\end{figure}

For simplicity, we calculate the radial drift timescale, $t_{\rm drift}$, for an irradiated disk in this section, but most of our conclusions hold true for an evolving disk as well. Figure \ref{fig:timescales} shows $t_{\rm des}$, $t_{\rm drift}$ and $t_{\rm gas, acc}$ as a function of particle size at three different locations in the disk, corresponding to the H$_2$O, CO$_2$ and CO snowlines in the static disk. As expected, micron-sized particles desorb on very short timescales of $\sim 1-1000$ years in the close vicinity of their respective snowlines, since the desorption rate depends exponentially on temperature and hence on disk location (see Equation \ref{eq:Rdes}).  On the other hand, their radial drift timescale exceeds the typical disk lifetime of a few Myr by several orders of magnitude due to their strong coupling with the gas. Thus for small particles in an irradiated disk, the snowline locations and the C/O ratio are the same as for a static disk (see Figure 1 from \citealt{oberg11}). This is not true for an evolving disk, however, where gas accretion causes even micron-sized particles to drift significantly before desorbing, as we show in section \ref{sec:snowlines}. At the other extreme, kilometer-sized particles are unaffected by gas drag and have long desorption timescales ($\gg$1 Myr ), and the snowline locations and C/O ratio remain unchanged in this case as well. This is true for both irradiated and evolving disks, since large planetesimals are decoupled from the gas and hence unaffected by gas accretion onto the host star. 

In the particle size regime for which (1) $t_{\rm drift} \lesssim t_{\rm des} \lesssim t_{\rm d}$ ($t_{\rm d}=3$ Myr is the disk lifetime), i.e. for $\sim$$0.5$ cm $\lesssim s \lesssim$ $1000$ cm, or (2) $t_{\rm gas, acc} \lesssim t_{\rm des} \lesssim t_{\rm d}$, i.e. for $\sim$$0.1$ cm $\lesssim s \lesssim$ $10$ cm, radial drift or gas accretion (or both) are faster than thermal desorption, which is of particular interest for our purposes. We note that $t_{\rm gas, acc}<t_{\rm d}$ always holds true. Particles of sizes that satisfy the requirements above will drift significantly due to radial drift or gas accretion before desorbing, thus moving the H$_2$O, CO$_2$ and CO snowlines closer towards the central star and changing the C/O ratio throughout the disk. We quantify these effects in sections \ref{sec:snowlines} and \ref{sec:COratio}.


\section{Snowline Locations}
\label{sec:snowlines}

\begin{figure*}[tb]
\centering
\includegraphics[width=0.7\textwidth]{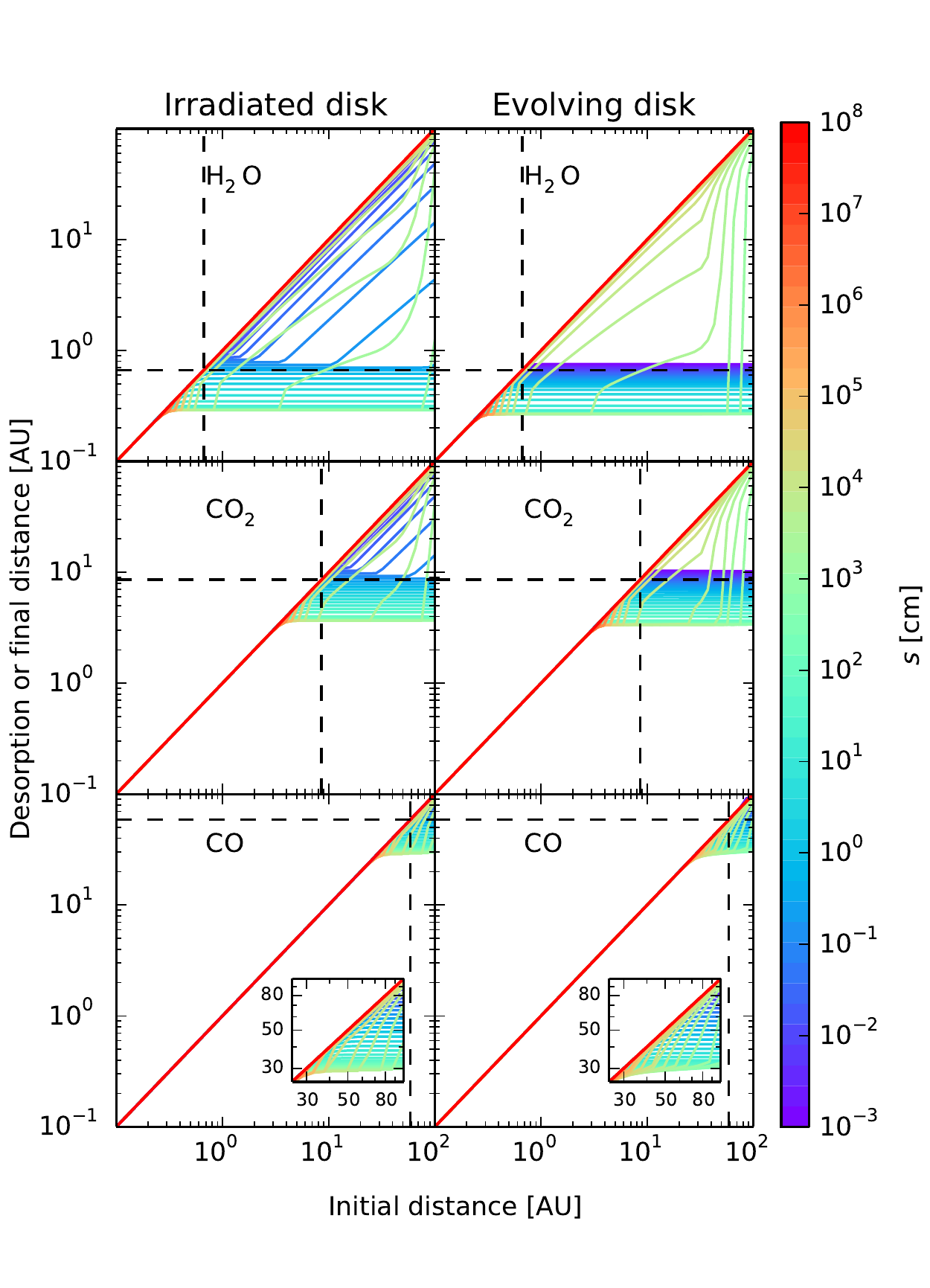}
\caption{Desorption distance (if a grain fully desorbs; horizontal lines) or final distance (if a grain does not fully desorb; diagonal lines for particles that do not drift and non-horizontal, non-diagonal lines for particles that drift), as a function of a particle's initial location in the disk, for a range of particle sizes, and for both an irradiated disk (left panels) and an evolving disk (right panels). The desorption distance is calculated for particles composed of H$_2$O (top panels), CO$_2$ (middle panels) and CO (bottom panels). The desorption distance for a static disk is shown for comparison (dashed vertical and horizontal lines). The particle size increases from $10^{-3}$ cm to $10^8$ cm as indicated by the color bar. For a particle of a given initial size that entirely desorbs during $t_{\rm d}=3$ Myr, the desorption distance is the same regardless of the particle's initial location.} 
\label{fig:snowlines}
\end{figure*}

\begin{figure*}[t!]
\centering
\includegraphics[width=0.7\textwidth]{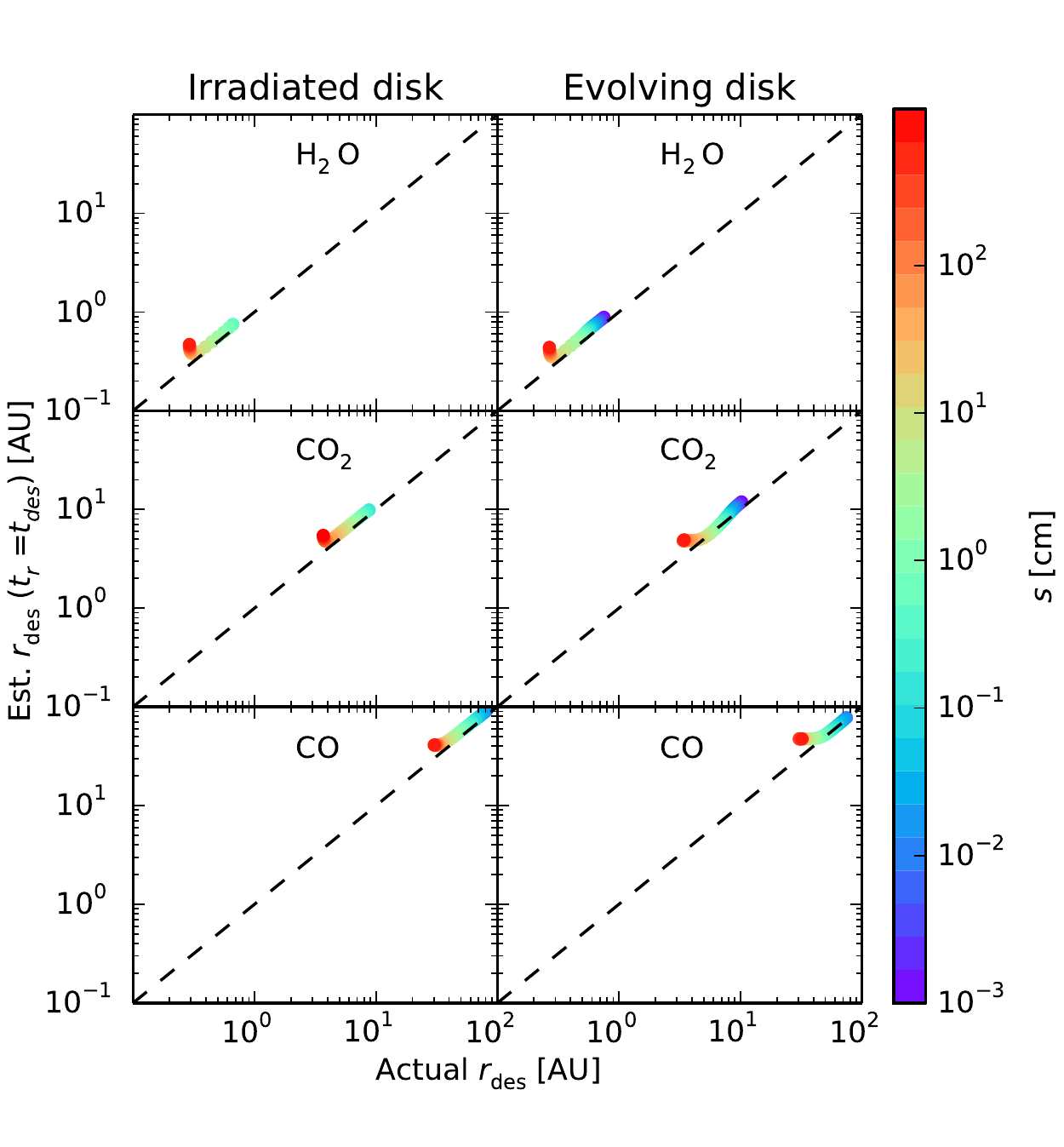}
\caption{Desorption distance estimated from analytic calculations (see text) as a function of the desorption distance calculated numerically, for the range of particle sizes that desorb at a fixed distance regardless of their initial location (see Figure \ref{fig:snowlines} and text). The estimate is performed for an irradiated disk (left panels) and an evolving disk (right panels).  The particles are composed of H$_2$O (top panels), CO$_2$ (middle panels) and CO (bottom panels). The analytic approximation is in good agreement with the numerical result for most cases, with the exception of larger particles, $s \gtrsim 10$ cm (see text).}
\label{fig:an_vs_actual}
\end{figure*}

In this section we use the model described in section \ref{sec:model} to quantify the effects of radial drift (irradiated disk) or radial drift and gas accretion (evolving disk) on the snowline location, for dust particles of different sizes composed of either H$_2$O, CO$_2$ or CO. Specifically, we determine a particle's final location (i.e., where the particle either fully desorbs or remains at its initial size due to having a desorption timescale longer than the time at which we stop the simulation) as a function of its initial position in the disk, after the gas disk has dissipated. The disk lifetime, $t_{\rm d}$, is particularly relevant since the timescale for giant planet formation must be less than or equal to $t_{\rm d}$. The snowline locations at $t=t_{\rm d}$ throughout the protoplanetary disk determine the disk C/O ratio in gas at this time, and thus the C/O ratio in giant planet atmospheres that have formed \textit{in situ}, before planetesimal accretion or core dredging.

For each species $x$, we determine the final location in the disk of a particle of initial size $s_0$ by solving the following system of coupled differential equations:

\begin{subeqnarray}
\label{eq:ddt}
\frac{ds}{dt} &= & - \frac{3 \mu_x m_{\rm p}}{\rho_{\rm s}} N_x R_{\rm des, x}  \slabel{eq:dsdt} \\
\frac{dr}{dt} &=& \dot{r} \slabel{eq:drdt},
\end{subeqnarray}
where the desorption rate $R_{\rm des, x}$ for each particle type (i.e., composed of H$_2$O, CO$_2$ or CO) is evaluated at $T=T(r)$, and the radial drift velocity $\dot{r}$ is given by Equation (\ref{eq:rdotpas}) for an irradiated disk and Equation (\ref{eq:rdotact}) for an evolving disk. Equations (\ref{eq:dsdt}) and (\ref{eq:drdt}) describe the coupled desorption and radial drift, and can be derived straightforwardly from Equation (\ref{eq:tdes}). Our initial conditions are $s(t_0)=s_0$ and $r(t_0)=r_0$, where $t_0$ is the initial time at which we start the integration and $r_0$ is the initial location of the particle.  We choose $t_0=1$ year, but our result is independent on the initial integration time as long as $t_0 \ll t_{\rm d}$. The desorption timescale $t_{\rm des}$ will then satisfy $s(t_{\rm des})=0$, from which we can determine the desorption distance $r_{\rm des}=r(t_{\rm des})$.



We define the final position of a grain as the disk location it has reached after $t_{\rm d}=3$ Myr, or the radius at which it completely desorbs if that happens after a time shorter than 3 Myr.  Figure \ref{fig:snowlines} shows our results for H$_2$O, CO$_2$ and CO particles, for both an irradiated and an evolving disk. We do not show the results for the viscous disk as they would complicate the plot without adding any qualitative insight ---  the results for the viscous disk are quantitatively similar with those of the evolving disk for the CO$_2$ and CO particles, but they are different for the H$_2$O grains, since accretion heating will push the H$_2$O snowline outwards (see Section \ref{sec:COratio}). We also show the static snowlines for comparison, which are calculated by balancing adsorption and desorption \citep{hollenbach09}. 
Kilometer-sized bodies do not drift or desorb during the disk lifetime neither for an irradiated nor for an evolving disk. Similarly, micron- to mm-sized particles in the irradiated disk do not drift and only 
desorb if 
they are located inside the static snowlines. 
In an evolving disk, however, micron-to mm-sized grains do drift significantly since they move at the same velocity as the accreting gas. For $0.5$ cm $\lesssim s_0 \lesssim$ 700 cm in an irradiated disk and $0.001$ cm $\lesssim s_0 \lesssim$ 700 cm in an evolving disk, we notice that particles of initial size $s_0$ desorb at a 
particle size dependent radius $r_{\rm des}$ regardless of their original location in the disk. In fact, the only grains that will both drift and evaporate are those that reach their fixed final location (represented by the horizontal curves in Figure \ref{fig:snowlines}) within the disk lifetime. We show in section \ref{sec:COratio} that this result is essential in determining the C/O ratio throughout the disk for different particle sizes. 

Another interesting feature of Figure \ref{fig:snowlines} is that particles above a certain size ($\sim$7 m for our choice of parameters) all desorb at the same distance. This is due to the fact that once the large bodies pass the static snowline, they first lose mass, thus eventually following the same evolutionary track as the meter-sized bodies and evaporating at the same location.

\begin{figure*}[t!]
\centering
\includegraphics[width=0.8\textwidth]{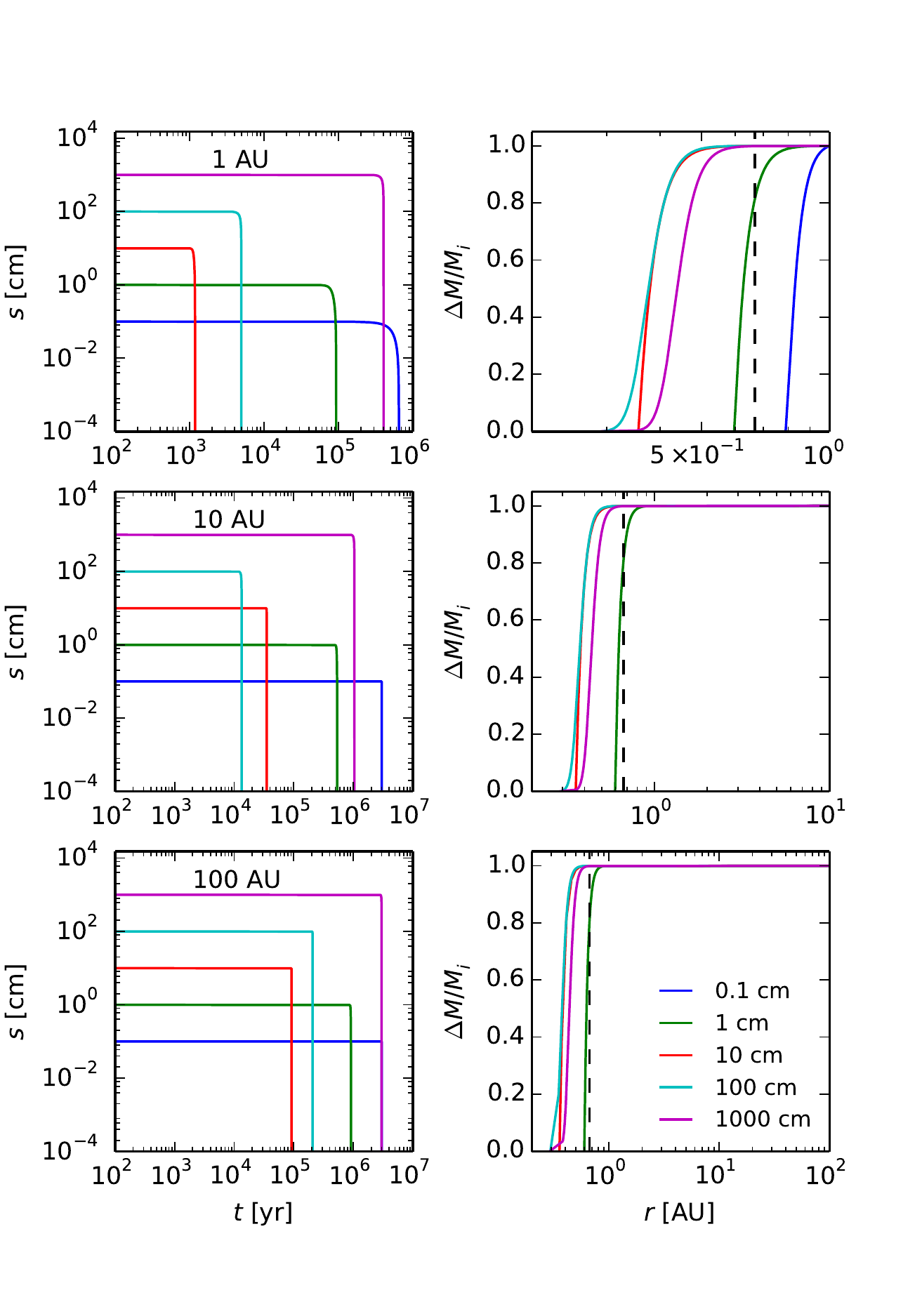}
\caption{Left panels: size of desorbing H$_2$O particles as a function of time, for different initial particle sizes and for three initial locations in an irradiated disk: 1 AU (top left), 10 AU (middle left) and 100 AU (bottom left). Particles desorb almost instantaneously. Right panel: fractional mass of the desorbing particles as a function of the particle's location as it drifts, for different initial particle sizes, and at the same initial locations presented in the left panel. Particles lose most of their mass very close to the distance at which they fully desorb. The static H$_2$O snowline is shown for reference (dashed vertical lines).}
\label{fig:s_t_a}
\end{figure*}

Intuitively, this fixed $r_{\rm des}$ should be the location in the disk for which $t_{\rm drift} \sim t_{\rm des}$, given an initial particle size. We can calculate this location analytically by equating Equations (\ref{eq:tdes}) and (\ref{eq:tdrift}) and solving for $r=r_{\rm des}(s)$ for a given particle size $s$. Figure \ref{fig:an_vs_actual} shows $r_{\rm des}$ calculated analytically using the prescription above as a function of the actual desorption distance calculated numerically. We display this result for the range of particle sizes that desorb at a fixed distance in an irradiated and an evolving disk (see Figure \ref{fig:snowlines}). We notice that the analytic approximation accurately reproduces the numerical result for most cases of interest, but it deviates for particles larger than $s \gtrsim10$ cm. For small particles with $\tau_{\rm s} \ll 1$, $t_{\rm drift}$ is a power-law in $r$ (for our parameters, $t_{\rm drift} \propto r^{-1/14}$ for the irradiated disk in the Epstein drag regime), and the Equation set (\ref{eq:ddt}) has an explicit analytic solution (see \App{app:tdriftan}). Once particles are large enough so that $\tau_{\rm s} \sim 1$, $t_{\rm drift}$ has a more complicated dependence on $r$ (see Equation \ref{eq:ts}), and the coupled drift-desorption differential equations have to be integrated numerically to obtain an accurate result.


Given $r_{\rm des}$, we need to only calculate the distance over which particles desorb to determine the location of a snowline. Figure \ref{fig:s_t_a}, left panels, shows the size evolution with time for H$_2$O particles of various initial sizes, starting at three different initial locations in an irradiated disk. Once solid H$_2$O particles begin to evaporate, they do so almost instantly for all explored particle sizes and initial locations. 
The right panels of Figure \ref{fig:s_t_a} show that the drifting grains lose most of their mass in a very narrow distance range; moreover, this distance is the same for a given initial particle size, no matter where the particle started drifting at the time $t_0$ when the simulation is started. Figure \ref{fig:s_t_a} thus demonstrates that solid particles that drift and fully desorb during the lifetime of the protoplanetary disk do so (1) instantaneously, and (2) at a fixed stellocentric distance, regardless of their initial location in the disk. 
It follows that the H$_2$O, CO$_2$ and CO snowlines are fixed for a given initial particle size and disk model. Both of these conclusions remain valid for an evolving and a viscous disk, as well as for particles composed of CO$_2$ or CO, but the snowline locations will vary between the three disks for a given initial particle size (see Section \ref{sec:COratio}). If we do not take into account the time dependence of the mass accretion rate and stellar luminosity (see Section \ref{sec:disk}), the C/O ratio will then only depend on disk properties, grain size, and the abundance of H$_2$O, CO$_2$ and CO relative to the H$_2$ abundance in the disk midplane, and {\it not} directly on the disk age when only considering drift, accretion and desorption.  




\section{C/O Ratio Estimates}
\label{sec:COratio}

\begin{figure}[h!]
\centering
\includegraphics[width=0.55\textwidth]{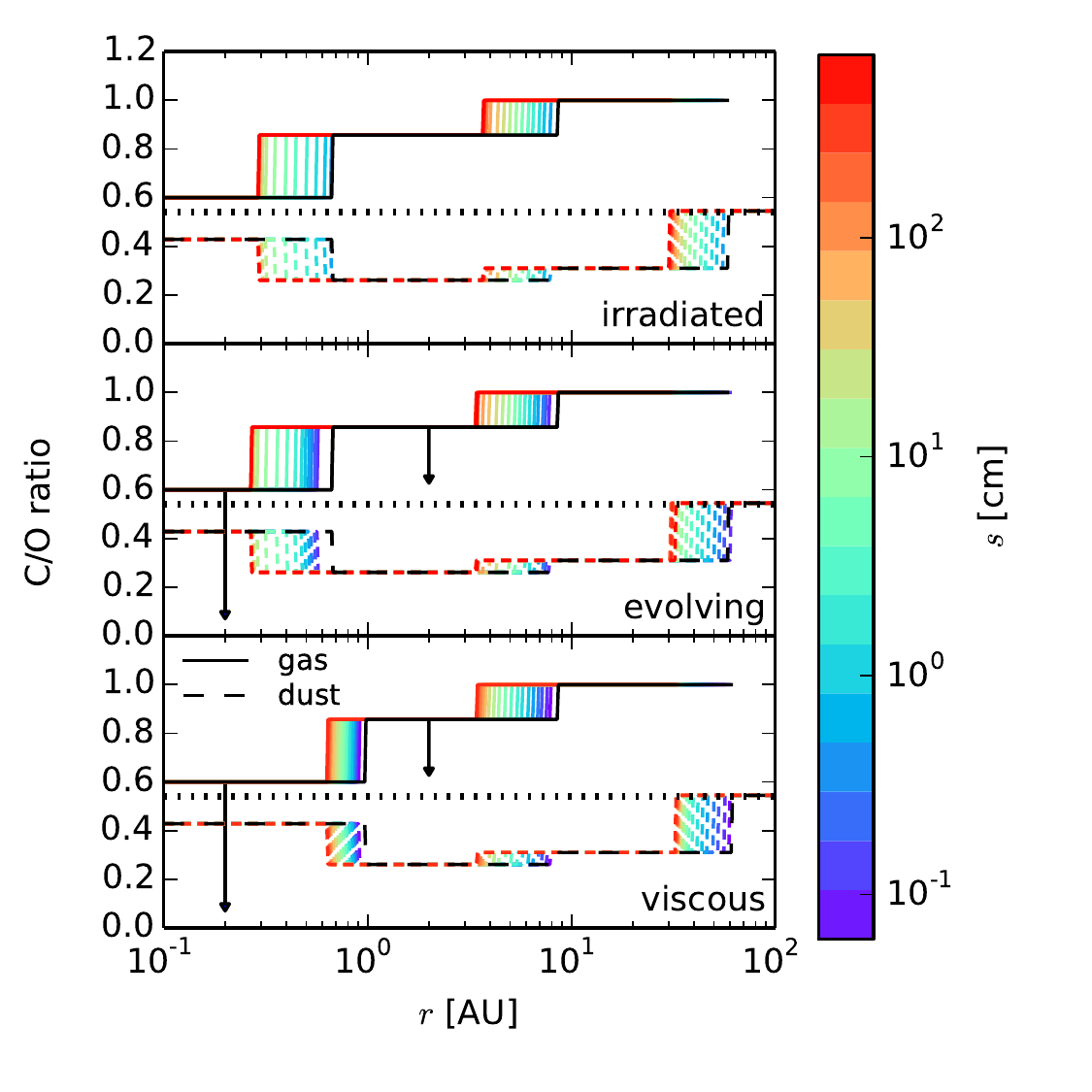}
\caption{Estimated C/O ratio in gas (solid lines) and in dust (dashed lines) for an irradiated disk (top panel), an evolving disk (middle panel) and a viscous disk (bottom panel).
The particle size increases from $\sim$0.05 cm to $\sim$700 cm as indicated by the color bar. The horizontal dotted line represents the stellar value of 0.54. The black lines represent the C/O ratio in gas (solid black line) and dust (dashed black line) for a static disk, with the temperature profile given by Equation (\ref{eq:diskT}) for the top two panels and by Equation (\ref{eq:activeT}) for the bottom panel. 
For both the evolving and the viscous disk, the movement of desorbed CO$_2$ gas inside the CO$_2$ snowline, and of desorbed CO$_2$ and H$_2$O gas inside the H$_2$O snowline due to gas accretion will increase the amount of oxygen gas inside the respective snowlines and thus reduce the gas C/O ratio, as shown by the arrows.}
\label{fig:CO_ratio}
\end{figure}

Given our results in Section \ref{sec:snowlines}, a disk's C/O ratio in mainly affected by the snowline location for the particle size housing the most mass in ice.
Realistic grain size distributions in disks are dominated by large grains (e.g., \citealt{dalessio01}, \citealt{birnstiel12}). In Figure \ref{fig:CO_ratio}, we display the H$_2$O, CO$_2$, and CO snowline locations as a function of particle size for disks with static chemistry that experience radial drift of solids and gas accretion onto the central star.  The minimum snowline distance for a disk is given by the curve corresponding to the maximum particle size it hosts.  For grains that have grown to radii larger than $\sim$7 m and are able to drift and desorb, the $\sim$7 m snowline applies (see Section 3).  

Drift and gas accretion affect the C/O ratio in a disk both because they move the snowline locations of the main C and O carriers  and because they cause solids and gas---which contain different proportions of C and O---to move inward at different rates.  As shown in Section 3, the snowline locations depend on disk age only indirectly, through changes in disk properties and grain size.  The C/O ratio is a function of the locations of the snowlines and the abundances of H$_2$O, CO$_2$, and CO relative to the H$_2$ abundance in the disk midplane.   These abundances evolve over time as solids and gas move inward at different rates.

Figure \ref{fig:CO_ratio} shows the estimated C/O ratio in gas and dust as a function of semimajor axis for an irradiated disk, an evolving disk, and a viscous disk, under the simplifying assumption that the abundance relative to hydrogen for each volatile is fixed, so that drift and accretion affect only the locations of the snowlines. We use the relative number densities of C and O in their different molecular forms (H$_2$O, CO$_2$ and CO) from Table 1 of \citet{oberg11}.  Snowline locations correspond to $r_{\rm des}$ in Figure \ref{fig:snowlines}, representing the location at which particles desorb in the absence of readsorption.  The C/O ratio for a static disk, where desorption and readsorption balance \citep{hollenbach09}, is shown as a guideline. We note that the true snowline for particles with $r_{\rm des}$ outside the static snowline is the static snowline itself---thus only particles with initial sizes larger than $\sim$0.05 cm are plotted in the three panels, as particles that form snowlines at larger distances (cf. Figure \ref{fig:snowlines}) are not true snowlines.

Before discussing the quantitative aspects of this plot, it is essential to acknowledge that our estimates for the C/O ratios in the evolving and viscous disks ignore the movement of the desorbed ices with the accreting gas---the relative fluxes of the volatiles in gaseous and solid form will affect the relative abundance of C and O in gas and dust throughout the disk. As demonstrated in Figure \ref{fig:s_t_a}, this will not affect the snowline locations for particles of a given size, but will change the shape of the C/O curves in between the various snowlines. For example, for the disk parameters and particle sizes displayed in Figure \ref{fig:CO_ratio}, water molecules in solid particles drift up to $\sim$1000 times faster across the H$_2$O snowline than do molecules of CO and CO$_2$ vapor that are entrained in the accreting gas.  This differential inward motion will result in an increased oxygen gas abundance inside the H$_2$O snowline, and thus a (in some cases much) lower gaseous C/O ratio in this region. Conversely, oxygen gas inside the water snowline will be depleted compared to the static disk if H$_2$O particles grow to planetesimal size and stall their migration between the H$_2$O and CO$_2$ snowlines, leaving only gaseous CO and CO$_2$ to accrete inward.  Growth of large planetesimals can therefore increase the C/O ratio in the inner disk.

Figure \ref{fig:CO_ratio} plots snowline curves for particle sizes $\sim0.5$ cm $\lesssim $s$ \lesssim$ 7m.  In the outermost disk, H$_2$O, CO$_2$, and CO all solidify.  Hence, relative drift across the CO snowline can alter only the abundances of volatiles between the CO$_2$ and CO snowlines, but not the C/O ratio in this region.  Interior to the CO$_2$ snowline, however, relative drift is important. 
We have found that the largest drifting particles in our model ($\sim$7 m) drift faster than the gas at both the H$_2$O and CO$_2$ snowlines. We thus conclude that the C/O ratio interior to the H$_2$O and CO$_2$ snowlines in our evolving and viscous disks will be lower than in the static disk, due to the additional oxygen added to the gas by desorbing H$_2$O and CO$_2$.  For these particle sizes, our calculated C/O ratio is an upper limit,  as indicated by the arrows in Figure \ref{fig:CO_ratio}.

Fundamentally, the elevated C/O ratios interior to the static H$_2$O and CO$_2$ snowlines are simply caused by the inward movement of the snowlines due to radial drift and gas accretion. Qualitatively, this scenario should be robust to changes in total abundances throughout the disk --- for example, at the dynamic (non-static) CO$_2$ snowlines, the rapid return of CO$_2$ into gas-phase during CO$_2$ desorption will reduce the C/O ratio interior to the CO$_2$ dynamic snowline, while no major change in gas-phase composition, and therefore C/O ratio, is expected between the static and dynamic snowlines.

As noted earlier, we assume that the total (ice and gas) abundance of each volatile is the same at every radius after ices have migrated. This is a good approximation for the irradiated disk, given that this model, by definition, presents a constant influx of particles at any given radius while the gas is static, and thus the ice and gas surface density remain constant. For the evolving disk, the gas-phase C/O ratio may decrease everywhere interior to the H$_2$O and CO$_2$ snowlines due to the decline in the surface density of solids with time at any given radius. For the viscous disk, the solid abundances at a fixed radius are constant, given that this model is not time-dependent, but the gas-to-solid ratio is not constant, which can result in a substantially lower C/O ratio interior to the H$_2$O and CO$_2$ snowlines compared to the static case (as indicated by the arrows in Figure \ref{fig:CO_ratio}). The main goal of Figure \ref{fig:CO_ratio}, however, is to show the different snowline radii in static and non-static disks, and therefore the locations in the disk where the gas-phase C/O ratio is reduced or increased, rather than provide a quantitative estimate of the magnitude of the C/O increase or decrease.

The snowline locations in these disks exhibit several interesting features. For the irradiated disk, only grains larger than $\sim$0.5 cm drift, desorb and thus move the snowline compared to the static disk. In contrast, even $\sim$micron-sized grains drift and desorb for the evolving disk, since they flow towards the host star together with the accreting gas. For the same particle size, the snowline locations are slightly closer to the central star in the evolving disk, due to the fact that the accreting gas adds an additional component to the drift velocity of the solids (cf. Equation \ref{eq:rdotact}). The addition of accretional heating in the viscous disk moves the H$_2$O snowline outwards. This is due to the fact that accretional heating dominates in the inner disk, where high temperatures cause the grains to evaporate further away from the star. Once $r\gtrsim1-2$  AU, stellar irradiation dominates the thermal evolution of the disk, and therefore the CO$_2$ and CO snowlines locations are the same as in the evolving and viscous disks. 

Perhaps the most interesting feature is the fact that the snowlines are pushed inwards as the grain size increases. While the plot only shows the snowlines and C/O ratio for particle sizes up to $\sim$7 m, we have found that 
bodies larger than $\sim$7 m evaporate at the same location as the meter-sized planetesimals (see Section \ref{sec:snowlines}). However, the contribution of kilometer-sized bodies to the snowline location is modest, since they only drift if they are located very close to the snowline.
Thus the innermost snowlines (depicted in red in Figure \ref{fig:CO_ratio}) set the limit on how close in the H$_2$O, CO$_2$ and CO snowlines can be pushed due to radial drift and gas accretion on to the host star. For a grain size distribution with a maximum particle size different than our model, one can pick out the appropriate minimum snowline locations from this plot.


For our choice of parameters, the minimum snowline radii are: $r_{\rm{H_2O}} \approx 0.3$ AU for the irradiated disk, $r_{\rm{H_2O}} \approx 0.26$ AU for the evolving disk and $r_{\rm{H_2O}} \approx 0.63$ AU for the viscous disk; $r_{\rm{CO_2}} \approx 3.7$ AU for the irradiated disk, $r_{\rm{CO_2}} \approx 3.4$ AU for both the evolving and the viscous disks; $r_{\rm{CO}} \approx 30$ AU for the irradiated and both the evolving and the viscous disks. For comparison, $r_{\rm{H_2O}} \approx 0.67$ AU\footnote{For the viscous disk, we calculated the static snowline location using the same temperature profile as that of the viscous disk, for consistency purposes. Thus $r_{\rm{H_2O}} \approx 0.98$ AU for the static disk in this scenario.}, $r_{\rm{CO_2}} \approx 8.6$ AU and $r_{\rm{CO}} \approx 59$ AU for the static disk. For the viscous disk model, which is the most realistic, radial drift and gas accretion push the snowline locations inwards by up to $\sim$$40$ \% for H$_2$O, by up to  $\sim$$60$ \% for CO$_2$, and by up to $\sim$$50$ \% for CO.  We note that the H$_2$O snowline in all disks is significantly closer to the host star compared with Solar system models, which place the H$_2$O snowline between $\sim$$2.7$ to $\sim$$3.1$ AU (\citealt{hayashi81}, \citealt{podolak04}, \citealt{martin12}). This is partially because we choose a colder disk model, as well as the fact that gas accretion rates decrease over time, moving the snowline location inwards (see also \citealt{garaud07} and Section \ref{sec:neglected}). \citet{min11} find that the location of the H$_2$O snowline is highly sensitive to the gas mass accretion rate $\dot{M}$ (equal to $10^{-8} M_{\odot}$ yr$^{-1}$ in our model) and the dust opacity $\kappa$ (Equation \ref{eq:opacity}). Higher values of $\dot{M}$ and $\kappa$ would increase the accretional component of the disk temperature (cf. Equation \ref{eq:Tdact}), which would push the H$_2$O snowline in the viscous disk outwards to match the Solar system snowline. At the same time, the snowline location in Solar type stars may be as close as $\sim$1 AU \citep{mulders15}, further in than the H$_2$O snowline in our Solar system.

 Observations of the CO snowline in TW Hya \citep{qi13} have found its location at a disk midplane temperature of 17 K (at 30 AU for the TW Hya specific temperature profile). The inferred  desorption temperature corresponds to the CO desorption temperature in a static disk, or to desorption from very small grains in an evolving disk, i.e. from grains that are too small to drift substantially. This suggests that the outer TW Hya disk is dominated by small grains, since larger particles would push the snowline location inwards, and therefore to higher desorption temperatures. This may seem contradictory to observations of grain growth in disks in general and in TW Hya in particular \citep{wilner00}. However, recent observations have revealed that grain growth is concentrated to the inner disk \citep{perez12} and outer disk snowlines may therefore be close to the ones expected in a static disk.

\section{Discussion}
\label{sec:discussion}

\begin{figure*}[t!]
\centering
\includegraphics[width=\textwidth]{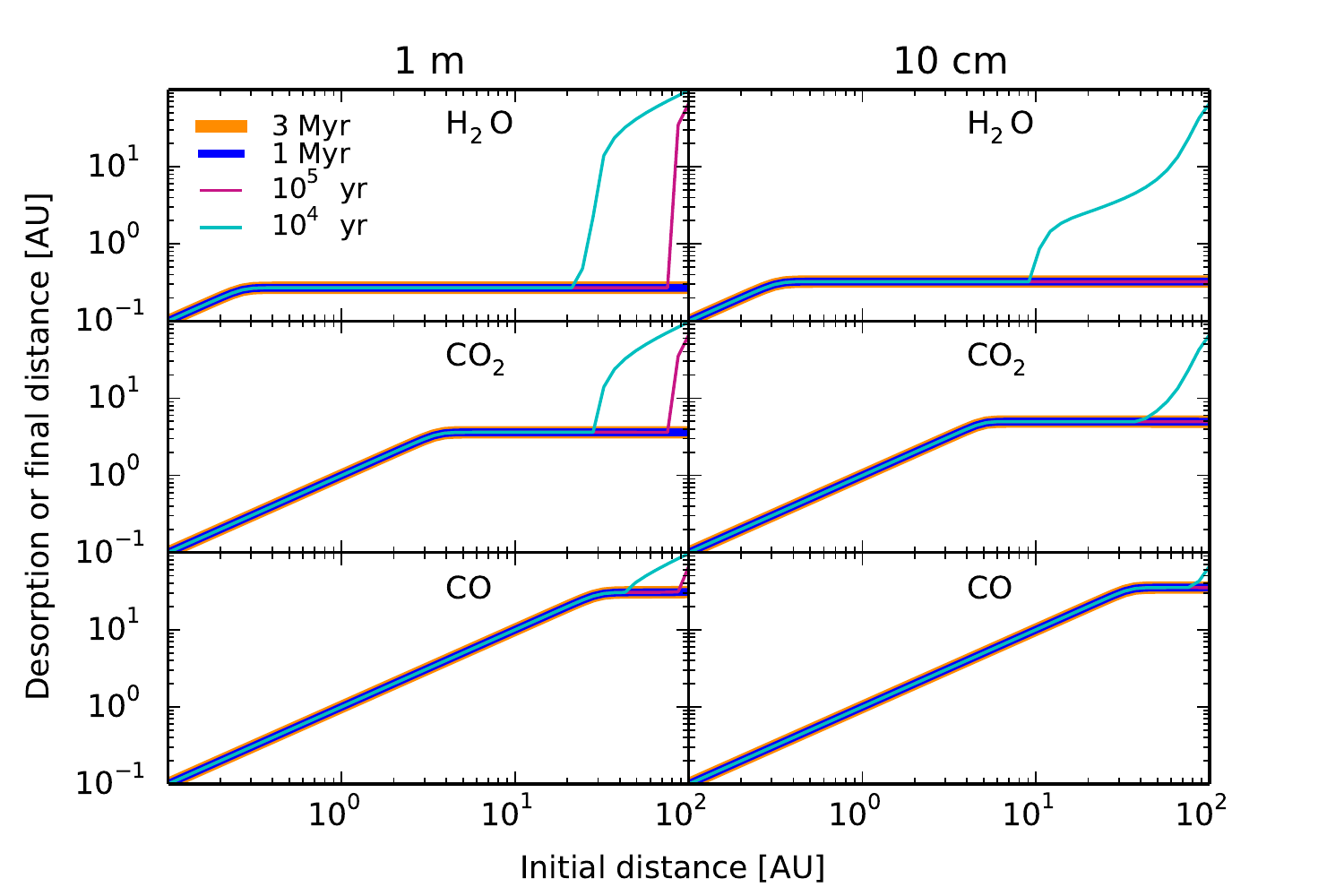}
\caption{Desorption or final distance as a function of initial position in the disk for particles of initial size $s_0=1$ m (left panels) and $s_0=10$ cm (right panels), for grains composed of H$_2$O (top panels), CO$_2$ (middle panels) and CO (bottom panels). The evolution is shown at four representative timescales: $10^4$ yr (cyan curve), $10^5$ yr (purple curve), 1 Myr (blue curve), and 3 Myr, the disk lifetime (orange curve). For a given particle size, the desorption distance, and hence the H$_2$O, CO$_2$ and CO snowlines, have the same location regardless of the time at which the simulation is stopped.}
\label{fig:timeplots}
\end{figure*}

\begin{figure}[h!]
\centering
\includegraphics[width=0.5\textwidth]{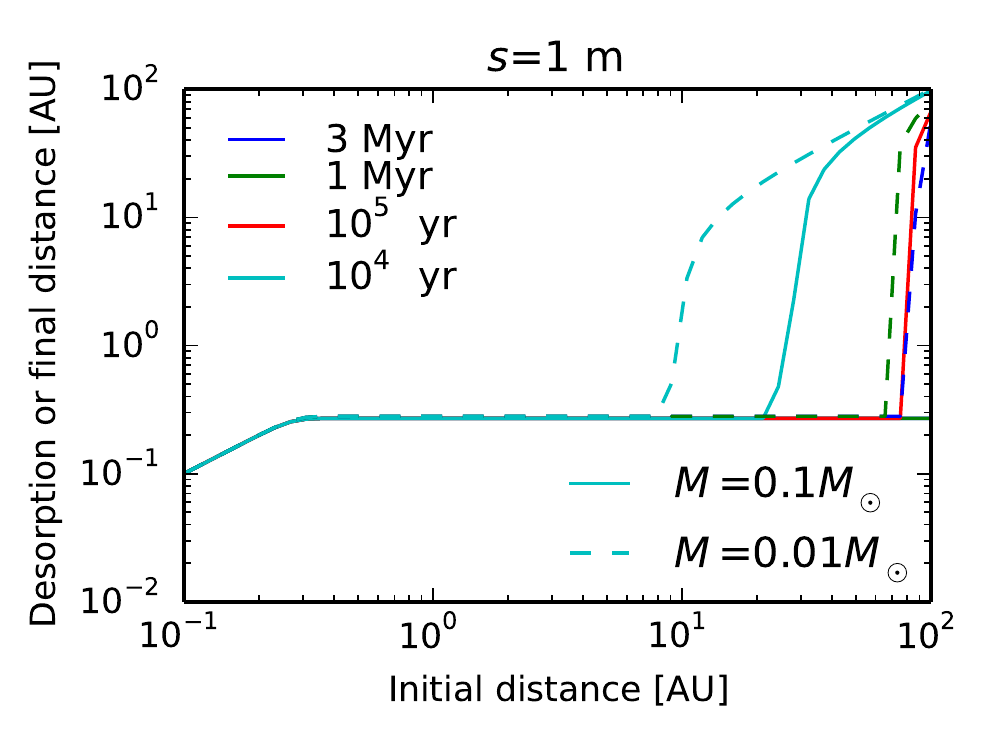}
\caption{Desorption or final distance as a function of initial position in the disk for H$_2$O particles of initial size of 1 m, for total disk masses $M=0.1 M_{\odot}$ (solid lines) and $M=0.01 M_{\odot}$ (dashed lines). The timescales at which we stop the simulations are $10^4$ yr (cyan curve), $10^5$ yr (red curve), 1 Myr (green curve) and 3 Myr (blue curve). A lower disk mass does not change the snowline location.}
\label{fig:varMd}
\end{figure}

\begin{figure}[h!]
\centering
\includegraphics[width=0.5\textwidth]{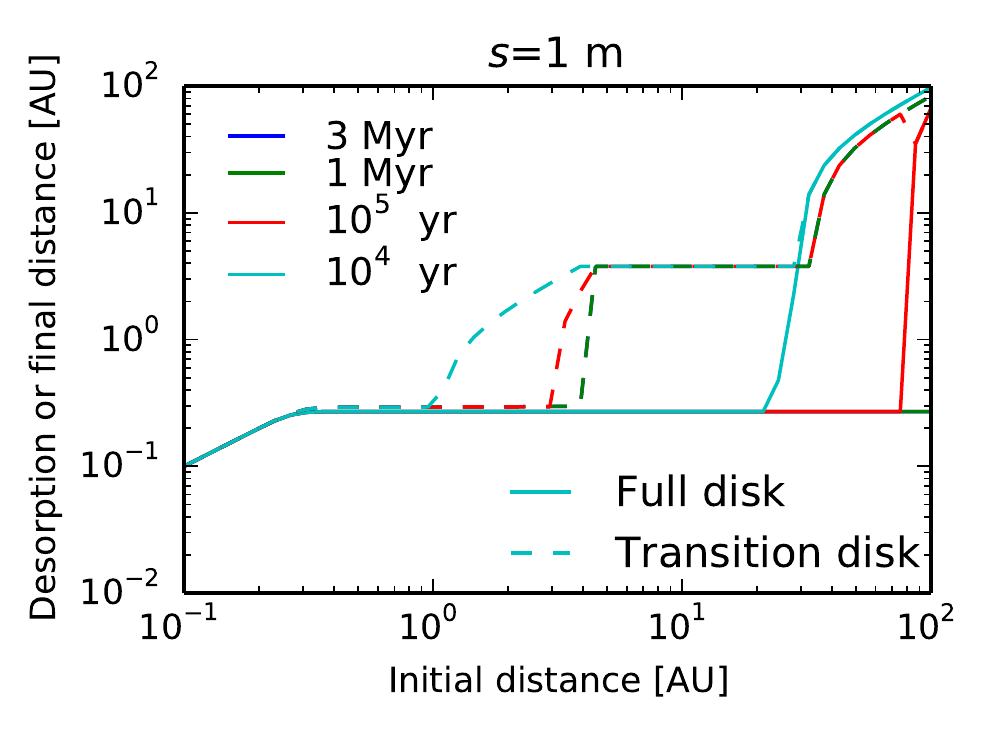}
\caption{Desorption or final distance as a function of initial position in the disk for H$_2$O particles of initial size of 1 m, for our fiducial disk (solid lines) and for a transition disk with an inner cavity at $r_0=4$ AU (dashed lines). The timescales of the simulations and their color code are the same as in Figure \ref{fig:varMd}. Particles that start inside the cavity drift towards the original snowline, while particles that start outside the gap stop shortly after crossing the gap edge, due to being trapped in a pressure maximum.}
\label{fig:cavity}
\end{figure}

\subsection{Generality of Results: Dependence on Disk Parameters}
\label{sec:incond}

In this section we investigate how variations in our fiducial parameters, the total disk mass, disk age, and disk structure, 
affect the calculated snowline locations and the C/O ratio. 
All previous results assumed a disk lifetime $t_{\rm d}=3$ Myr
, the typical disk lifetime and the expected time scale for giant planets to accrete their gaseous atmopsheres (e.g., \citealt{pollack96}, \citealt{piso14}). Some gas accretion may occur at earlier times, however,
before the core is fully formed (e.g., \citealt{rafikov06}). Recent models such as aerodynamic pebble accretion \citep{lambrechts12} suggest that rapid core growth on timescales of $10^5$ years is possible. 
The composition of giant planet atmospheres, and specifically their C/O ratio, can thus depend on the abundance of H$_2$O, CO$_2$ and CO in gas at earlier times than $t_{\rm d}$ in the disk evolution. 

Figure \ref{fig:timeplots} shows the particle desorption or final distance as a function of a particle's initial location in the disk, for ice particles 
of initial sizes of 10 cm and 1 m, composed of either H$_2$O, CO$_2$ or CO. These sizes are important since radial drift timescales are shortest for particles within this size range (see Figure \ref{fig:timescales}) --- these are the particles whose drift and desorption evolution should be most strongly affected by variations in disk conditions. We choose the evolving disk as a disk model, and we stop the simulations after $10^4$ yr, $10^5$ yr, 1 Myr and $t_{\rm d}=3$ Myr, respectively. 
The most important result of these plots is that particles of a given size always desorb at the same disk radii, the 3 Myr snowline, regardless of simulation stopping time.  Particles 
that start at large stellocentric distances do not desorb within the shorter timeframes, e.g. $10^4$ or $10^5$ years, but they do evaporate at a fixed radius if their initial location is closer to the host star. While the amount of material that moves through the disk changes with time, the radius at which particles desorb and 
the snowline locations are thus independent of the time elapsed, and 
our results for the snowline locations are 
valid throughout the time evolution of the protoplanetary disk. 

We choose as a fiducial model a total disk mass $M=0.1 M_{\odot}$. Observationally, disk masses span at least an order of magnitude around Solar type stars 
\citep{andrews13}. 
We thus explore the effect of disk mass on the location of snowlines. Figure \ref{fig:varMd} shows the desorption or final distance as a function on the initial location of a H$_2$O particle with initial size of 1 m, for two total disk masses: $M=0.1 M_{\odot}$, our fiducial model, and $M=0.01 M_{\odot}$. Similarly to Figure \ref{fig:timeplots}, we perform our calculations for an evolving disk. The simulations are stopped after the same timeframes as those in Figure \ref{fig:timeplots}. The location of the H$_2$O snowline is the same for both disks (the same holds true for the CO$_2$ and CO snowlines). The C/O ratio is thus insensitive to the choice of $M$. We note that our conclusions regarding the disk age and total mass are only valid if the snowline itself does not move with time or disk mass (see also Sections \ref{sec:disk} and \ref{sec:neglected}).


We also apply our evolving disk model to a transition disk, i.e. a protoplanetary disk with an inner cavity significantly depleted of gas. We choose a disk with an inner gap of radius $r_0=4$ AU, consistent with observations of TW Hya \citep{zhang13}, and with the gas surface density in the gap reduced by a factor of 1000. Figure \ref{fig:cavity} shows the desorption or final distance for a H$_2$O particle of initial size of 1 m, with the simulation stopped at the same timescales as in Figures \ref{fig:timeplots} and \ref{fig:varMd}. Particles that start at an initial distance interior to the gap drift towards the original snowline, while grains located exterior to the gap stop shortly after crossing the gap edge, due to the decrease in gas pressure inside the cavity, thus forming a snowline at $\sim$$3.8$ AU. This is qualitatively consistent with the observations of \citet{zhang13}, which show that the H$_2$O snowline is pushed outwards in a transition disk compared to a full disk. Our model framework is thus generally valid for more complicated disk structures as well.  


\subsection{Model Extensions}
\label{sec:neglected}

Our goals in this paper were (1) to gain a qualitative and quantitative understanding of the effect of radial drift and gas accretion onto the central star on snowline locations and the C/O ratio in disks, and (2) to obtain a limit on how close in the snowlines can be pushed due to drift and gas accretion. We have thus used a simplified model and out of necessity neglected potentially significant dynamical and chemical processes. In what follows, we discuss these limitations and their effects. We note that our future work will address some of these issues.

We summarize in Table 1 the potential physical and chemical processes occurring in disks and their effect on snowline locations compared to a static disk. For the sake of completeness, Table 1 also includes the processes addressed in this paper, i.e. radial drift and gas accretion. The neglected effects are discussed in more detail below.

\begin{enumerate}
\item \textbf{Particle growth.} While our model assumes a range of particle sizes, each size is considered initially fixed for a given grain before it drifts and desorbs, since we do not take into account particle coagulation. However, grain growth has been observed in protoplanetary disks (e.g., \citealt{ricci10}, \citealt{perez12}), as well as theoretically constrained (e.g., \citealt{birnstiel10}, \citealt{birnstiel12}). In Section \ref{sec:COratio} we have shown that larger grains move the snowline locations closer in, but those locations remain fixed above a certain particle size. Particle growth will thus initially push the snowlines inwards. This is consistent with particle growth models, which predict a maximum particle size often around or below the particle sizes that drift the fastest \citep{birnstiel12}. As the largest grains contain most of the solid mass, grain growth models should produce snowlines corresponding to our snowline location estimates for the largest grains in the particle size distribution. However, once the solids grow larger than km-sized and form planetesimals, they are no longer affected by drift or desorption, and the snowline reduces to that of a static disk. 

\item \textbf{Turbulent diffusion.} The radial drift model presented in Section \ref{sec:drift} only considers a laminar flow and thus ignores turbulence. However, the disk gas also experiences turbulent diffusion (e.g., \citealt{birnstiel12}, \citealt{alidib14}). Turbulence causes eddies and vertical mixing, which are likely to reduce the radial drift velocity of the solids (e.g., \citealt{youdin07}). Additionally, the flow of H$_2$O, CO$_2$ and CO vapor will diffuse radially. Back-diffusion across the snowline will change the shape of the snowline, as well as the C/O ratio in gas and dust both inside and outside of the snowline, due to the reduction of gas-phase volatile abundance interior to the snowline. 

\begin{table}[t!]
\caption{The effects of dynamical and chemical processes on snowline shapes and locations}
\begin{center}
\begin{tabular}{|l|l|}\hline
\textbf{Process} & \textbf{Effect} \\\hline
Radial drift & $\leftarrow$ \footnote{The arrows signify how a process affects the snowline: $\leftarrow$ means that the snowline is pushed closer to the host star compared to the static snowline, $\rightarrow$ means that the snowline is pushed further from the host star compared to the static snowline. The presence of both arrows means that the process may have both effects on the snowline location.} \\\hline
Gas accretion & $\leftarrow$ \footnote{Gas accretion pushes the snowlines inwards compared to the snowline locations in a static disk. However, accretional heating may push the snowline outwards compared to an evolving disk without accretion heating.} \\\hline
Particle growth & $\leftarrow$ \footnote{As stated in the main text, if particles grow to km-sizes and above, the snowline remains the same as that of a static disk.} \\\hline
Turbulent diffusion & $\rightarrow$ $\leftarrow$ \\\hline
Particle fragmentation & $\rightarrow$ $\leftarrow$ \\\hline
Grain morphology & $\rightarrow$ \\\hline
Particle composition & $\rightarrow$ $\leftarrow$ \\\hline
Disk gaps and holes & $\rightarrow$ \\\hline
Accretion rate evolution & $\rightarrow$ $\leftarrow$ \\\hline
Stellar luminosity evolution & $\leftarrow$ \\\hline
Non-static chemistry & $\rightarrow$ $\leftarrow$ \\\hline
\end{tabular}

\end{center}
\end{table}

\item \textbf{Particle fragmentation.} Frequent particle collisions in disks cause them to fragment (e.g., \citealt{birnstiel12}). The fragmentation of meter- to km-sized particles will move the snowlines outwards, as smaller particles desorb faster and further out from the host star (cf. Figures \ref{fig:snowlines} and \ref{fig:CO_ratio}). Large boulders, which neither drift nor desorb, may become e.g. meter-sized due to collisions and subsequent fragmentation, which will cause them to drift significantly before desorbing, pushing the snowlines inwards. Thus fragmentation can move the snowline locations in either radial direction --- specifically, fragmentation leads to a certain grain size distribution, and the largest particles in this size distribution are the ones that determine the position of the snowline. 

\item \textbf{Grain morphology.} Our model assumes that the ice particles are perfect, homogeneous spheres. However, this is not a very good approximation, since grain growth can be fractal rather than compact (\citealt{zsom10}, \citealt{okuzumi12}, \citealt{krijt15}). The inhomogeneity due to cracks in the grain structure will cause the particles to desorb faster. They will therefore drift less before evaporating and will move the snowlines less far inward.

\item \textbf{Particle composition.} The ice particles in our model are assumed to be fully made of either H$_2$O, CO$_2$ or CO. More realistically, grains may have a layered structure, such as an interior composed of non-volatile materials (e.g., sillicates) covered by an icy layer. The ice thus only constitutes a fraction of the total particle mass, which accelerates its desorption and pushes the snowlines outwards. The grains may also be composed of a mixture of H$_2$O, CO$_2$ and CO ices, which will increase the binding energies of the more volatiles species, moving 
the snowlines inwards. 


\item \textbf{Disk gaps and holes.} The snowline locations will be different for transition disks, which have inner cavities significantly depleted of gas (e.g., \citealt{espaillat12}, \citealt{vandermarel15}), or pre-transitional disks, which have a gap between an inner and outer full disk (e.g., \citealt{kraus11}). The decrease in gas pressure in these gaps or holes will reduce the particles' drift velocity close to the gap edge, thus slowing them down and pushing the snowline outwards.

\item \textbf{Accretion rate evolution.} Our viscous disk model assumes a constant mass accretion rate $\dot{M}$. However, $\dot{M}$ decreases over time, which lowers the accretional component of the disk temperature (Equation \ref{eq:Tdact}), thus pushing the snowline location inwards if the disk is optically thick (\citealt{garaud07}). Once $\dot{M}$ reaches low enough values for the snowline to become optically thin, the snowline location moves outwards \citep{garaud07}. During the giant planet formation stage of a few Myr, however, $\dot{M}$ steadily decreases with time \citep{chambers09}, which may result in the inward movement of the H$_2$O snowline by up to one order of magnitude, significantly larger than the inward movement caused by radial drift (cf. Section \ref{sec:COratio}). We thus acknowledge that the location of the H$_2$O snowline may be set by the mass accretion rate evolution rather than the drift of solids.



\item \textbf{Stellar luminosity evolution.} As the host star contracts during its pre-main sequence phase, its luminosity decreases, which reduces the disk temperature and pushes the snowline locations inwards. \citet{kennedy06} found that the snowline is unlikely to move significantly during the pre-main sequence phase for Solar type stars, but it may move inward by a factor of $\sim$$15-20$ for $M_* \sim 0.25 M_{\odot}$ due to the stellar contraction. 

\item \textbf{Time dependent chemistry.} As the goal of this paper was to explore only the dynamical effects on snowline locations and the C/O ratio in disks, we have assumed a simple, static chemical model. In reality, the chemistry in most of the disk is expected to be time-dependent.
In the inner disk, chemistry approaches equilibrium due to intense sources of ionizing radiation (e.g., \citealt{ilgner04}), while in the outer disk high energy radiation and cosmic rays are the key drivers of chemistry, which is no longer in equilibrium (e.g., \citealt{vandishoeck06}). A multitude of chemical evolution models have been developed (see references in \citealt{henning13}), many of which contain tens or hundreds of chemical reactions. Due to the complexity of these chemical models, most of them are decoupled from disk dynamics. The effect of disk chemistry on snowline locations, shape, time evolution, or the C/O ratio is therefore difficult to estimate.  


\end{enumerate}


\section{Summary}
\label{sec:summary}

We study the effect of radial drift of solids and viscous gas accretion onto the central star on the H$_2$O, CO$_2$ and CO snowline locations and the C/O ratio in a protoplanetary disk, assuming static chemistry. We develop a simplified model to describe the coupled drift-desorption process and determine the time evolution of particles of different sizes throughout the disk. We assume that the solid particles are perfect, homogeneous spheres, fully composed of either H$_2$O, CO$_2$ or CO. We apply our model to an irradiated disk, an evolving disk, and a viscous disk that also takes into account stellar irradiation. We determine the desorption or final location of drifting particles after a time equal to the disk lifetime, and use this result to set an inner limit for the location of the H$_2$O, CO$_2$ and CO snowlines. Our results can be summarized as follows:

\begin{enumerate}
\item Radial drift and gas accretion affect desorption and move the snowline locations inward compared to a static disk for particles with sizes $\sim$$0.5$ cm $\lesssim s \lesssim$ 7 m for an irradiated disk and $\sim$$0.001$ cm $\lesssim s \lesssim$ 7 m for an evolving disk. 

\item For our simplified model that does not account for the effects outlined in Section \ref{sec:neglected}, particles with sizes in the above range desorb almost instantaneously once desorption has begun, and at a fixed location in the disk that only depends on the particle size and the gas accretion rate. Thus for each particle size there is a fixed and uniquely determined H$_2$O, CO$_2$ or CO snowline. 

\item The results of our numerical simulation are in agreement with the analytic solution of the drift-desorption system of differential equations if the stopping time $\tau_{\rm s} \ll 1$. We present an explicit analytic solution for the desorption distance in this regime.

\item Since realistic grain size distributions are dominated in mass by the largest particles, the H$_2$O, CO$_2$ and CO snowlines are those created by the largest drifting particles in our model. This corresponds to the innermost snowlines that we determine. Our model thus sets a limit on how close to the central star the snowlines can be pushed by radial drift and gas accretion.

\item The snowline locations move inwards as the particle size increases; the innermost snowline is set by particles with initial size $s \sim 7$ m in our model --- bigger particles drift too slowly to make it further in before desorbing (see Section \ref{sec:snowlines}). Gas accretion causes even micron-sized particles to drift, desorb and move the snowline location compared to a static disk. A viscous disk that includes accretion heating moves the H$_2$O snowline outwards compared to an evolving disk, but has no effect on the CO$_2$ and CO snowline locations, for our particular choice of mass accretion rate $\dot{M}$ and midplane opacity $\kappa$. 

\item For our fiducial model, which considers particles with sizes between $10^{-3}$ and $10^8$ cm, the innermost H$_2$O, CO$_2$ and CO snowlines are located at 0.3 AU, 3.7 AU and 30 AU for an irradiated disk, 0.26 AU, 3.4 AU and 30 AU for an evolving disk, and 0.63 AU, 3.4 AU and 30 AU for a viscous disk with accretion heating. Compared to a static disk, radial drift and gas accretion move the snowlines by up to 60 \% for H$_2$O and CO$_2$, and by up to 50 \% for CO. For the viscous disk, however, which is the most realistic of the three models since it takes into account accretion heating, the H$_2$O snowline location moves inwards by up to 40 \%.

\item Our C/O estimates confirm the conclusions of \citet{oberg11} that the C/O ratio in gas may be enhanced compared to the stellar value throughout most of the disk, with the C/O ratio reaching its maximum value between the CO$_2$ and CO snowlines. 
We note, however, that our results for the C/O ratio do not take into account the radial movement of the desorbed ices with the accreting gas in the evolving and viscous disks, which may significantly decrease the C/O ratio in gas inside the H$_2$O and CO$_2$ snowlines. We plan to address this issue in a future paper.

\item For a constant gas mass accretion rate $\dot{M}$ and stellar luminosity $L_*$, the snowline locations are independent of the time at which we stop our simulation and of the total disk mass, as long as the disk midplane remains optically thick. 

\end{enumerate}

Our model does not address additional effects, such as gas diffusion, grain composition and morphology, or complex time-dependent chemical processes. Future work will address some of these dynamical and chemical processes, with the goal of obtaining more realistic results for the snowline locations, shapes and time evolution, and the resulting effect on the C/O ratio. 

\acknowledgements{We thank the referee for a helpful and detailed report. We also thank Sean Andrews for useful conversations and suggestions. T. B. is grateful for support from the NASA Origins of Solar Systems grant NNX12AJ04G and the Smithsonian Institution Pell Grant program.}

\appendix
\section{Steady-state active disk solution} \label{app:steadystate}

Following \citet{shakura73} and \citet{armitage10}, the steady-state solution for a geometrically thin, optically thick actively accreting disk with an $\alpha$-prescription for viscosity is governed by the following set of equations:

\begin{subeqnarray}
\label{eq:diskeq}
\nu & = & \alpha c H \slabel{eq:nueq} \\
c^2 & = & \frac{k_{\rm B} T_{\rm act}}{\mu m_{\rm p}} \slabel{eq:cdeq} \\
\rho & = & \frac{1}{\sqrt{2 \pi}} \frac{\Sigma}{H} \slabel{eq:rhoeq} \\
H & = & \frac{c}{\Omega_{k}} \slabel{eq:Heq} \\
T_{\rm act}^4 & = & \frac{3}{4} \tau T_{\rm surf}^4 \slabel{eq:Teq} \\
\tau & = & \frac{1}{2} \Sigma \kappa \slabel{eq:taueq} \\
\nu \Sigma & = & \frac{\dot{M}}{3 \pi} \slabel{eq:Mdot} \\
\sigma T_{\rm surf}^4 & = & \frac{9}{8} \nu \Sigma \Omega_{\rm k}^2 \slabel{eq:nusigeq} \\
\kappa & = & \kappa_0 T_{\rm act}^2 \slabel{eq:kappaeq},
\end{subeqnarray}
where $T_{\rm surf}$ is the surface temperature of the disk and the other quantities are defined in the main text. This is a system of nine equations with nine unknowns ($\nu$, $c$, $H$, $T_{\rm act}$, $\rho$, $\Sigma$, $\tau$, $T_{\rm surf}$, $\kappa$) that can be solved numerically once $\alpha$ and $\kappa_0$ are specified.

\section{Desorption distance analytic solution} \label{app:tdriftan}

For a particle of size $s$ that desorbs and satisfies $\tau_{\rm s} \ll 1$ ($\tau_{\rm s}$ is the dimensionless stopping time, defined in Section \ref{sec:drift}), we can derive an explicit analytic solution for the particle's desorption distance in an irradiated disk. For $\tau_{\rm s} \ll 1$, a particle is in the Epstein drag regime (see Equation \ref{eq:ts}) and its drift velocity $\dot{r}$ (Equation \ref{eq:rdotpas}) can be approximated as
\begin{equation}
\label{eq:rdotapp}
\dot{r} \approx -2 \eta \Omega_{\rm k} r \tau_{\rm s}.
\end{equation}
By using Equations (\ref{eq:tdes}) and (\ref{eq:tdrift}) and setting $t_{\rm drift}=t_{\rm des}$, we can express a particle's desorption distance as

\begin{equation}
\label{eq:tan}
r_{\rm des}=\Bigg(\frac{d}{q \,C} \,\, \mathcal{W}\Big[\frac{(B/A)^{-q/d} q\,C}{d}\Big]\Bigg)^{\frac{1}{q}},
\end{equation}
where $\mathcal{W}$ is the Lambert-W function, $q=3/7$ is the power-law coefficient in Equation (\ref{eq:diskT}), $d=-\frac{1}{2}+p-q$ with $p=1$ the power-law coefficient in Equation (\ref{eq:disksigma}), and 
\begin{subeqnarray}
A & = &  \frac{\rho_0}{\rho_{\rm s}} \frac{r_0^2}{s c_0} r_0^d \\ 
B & = & \frac{\rho_{\rm s} s}{3 \mu_{\rm x} N_{\rm x} \nu_{\rm x}} \\
C & = & \frac{E_{\rm x}}{T_0} r_0^{-q}, \\
\end{subeqnarray}
where $r_0=1$ AU, $\rho_0=\rho(r_0)$ and $c_0=c(r_0)$.

%

\if\bibinc n
\bibliography{refs}

\begin{thebibliography}{}
\expandafter\ifx\csname natexlab\endcsname\relax\def\natexlab#1{#1}\fi

\bibitem[{{Aikawa} {et~al.}(1996){Aikawa}, {Miyama}, {Nakano}, \&
  {Umebayashi}}]{aikawa96}
{Aikawa}, Y., {Miyama}, S.~M., {Nakano}, T., \& {Umebayashi}, T. 1996, \apj,
  467, 684

\bibitem[{{Ali-Dib} {et~al.}(2014){Ali-Dib}, {Mousis}, {Petit}, \&
  {Lunine}}]{alidib14}
{Ali-Dib}, M., {Mousis}, O., {Petit}, J.-M., \& {Lunine}, J.~I. 2014, \apj,
  785, 125

\bibitem[{{Andrews} {et~al.}(2013){Andrews}, {Rosenfeld}, {Kraus}, \&
  {Wilner}}]{andrews13}
{Andrews}, S.~M., {Rosenfeld}, K.~A., {Kraus}, A.~L., \& {Wilner}, D.~J. 2013,
  \apj, 771, 129

\bibitem[{{Andrews} {et~al.}(2010){Andrews}, {Wilner}, {Hughes}, {Qi}, \&
  {Dullemond}}]{andrews10}
{Andrews}, S.~M., {Wilner}, D.~J., {Hughes}, A.~M., {Qi}, C., \& {Dullemond},
  C.~P. 2010, \apj, 723, 1241

\bibitem[{{Armitage}(2010)}]{armitage10}
{Armitage}, P.~J. 2010, {Astrophysics of Planet Formation (Cambridge, UK:
  Cambridge University Press)}

\bibitem[{{Bell} \& {Lin}(1994)}]{bell94}
{Bell}, K.~R., \& {Lin}, D.~N.~C. 1994, \apj, 427, 987

\bibitem[{{Birnstiel} {et~al.}(2010){Birnstiel}, {Dullemond}, \&
  {Brauer}}]{birnstiel10}
{Birnstiel}, T., {Dullemond}, C.~P., \& {Brauer}, F. 2010, \aap, 513, A79

\bibitem[{{Birnstiel} {et~al.}(2012){Birnstiel}, {Klahr}, \&
  {Ercolano}}]{birnstiel12}
{Birnstiel}, T., {Klahr}, H., \& {Ercolano}, B. 2012, \aap, 539, A148

\bibitem[{{Chambers}(2009)}]{chambers09}
{Chambers}, J.~E. 2009, \apj, 705, 1206

\bibitem[{{Chiang} \& {Youdin}(2010)}]{chiang10}
{Chiang}, E., \& {Youdin}, A.~N. 2010, Annual Review of Earth and Planetary
  Sciences, 38, 493

\bibitem[{{Ciesla} \& {Cuzzi}(2006)}]{ciesla06}
{Ciesla}, F.~J., \& {Cuzzi}, J.~N. 2006, Icarus, 181, 178

\bibitem[{{Collings} {et~al.}(2004){Collings}, {Anderson}, {Chen}, {Dever},
  {Viti}, {Williams}, \& {McCoustra}}]{collings04}
{Collings}, M.~P., {Anderson}, M.~A., {Chen}, R., {et~al.} 2004, \mnras, 354,
  1133

\bibitem[{{Cuzzi} \& {Zahnle}(2004)}]{cuzzi04}
{Cuzzi}, J.~N., \& {Zahnle}, K.~J. 2004, \apj, 614, 490

\bibitem[{{D'Alessio} {et~al.}(2001){D'Alessio}, {Calvet}, \&
  {Hartmann}}]{dalessio01}
{D'Alessio}, P., {Calvet}, N., \& {Hartmann}, L. 2001, \apj, 553, 321

\bibitem[{{Espaillat} {et~al.}(2012){Espaillat}, {Ingleby}, {Hern{\'a}ndez},
  {Furlan}, {D'Alessio}, {Calvet}, {Andrews}, {Muzerolle}, {Qi}, \&
  {Wilner}}]{espaillat12}
{Espaillat}, C., {Ingleby}, L., {Hern{\'a}ndez}, J., {et~al.} 2012, \apj, 747,
  103

\bibitem[{{Frank} {et~al.}(2002){Frank}, {King}, \& {Raine}}]{fkr02}
{Frank}, J., {King}, A., \& {Raine}, D.~J. 2002, {Accretion Power in
  Astrophysics: Third Edition}

\bibitem[{{Fraser} {et~al.}(2001){Fraser}, {Collings}, {McCoustra}, \&
  {Williams}}]{fraser01}
{Fraser}, H.~J., {Collings}, M.~P., {McCoustra}, M.~R.~S., \& {Williams}, D.~A.
  2001, \mnras, 327, 1165

\bibitem[{{Garaud} \& {Lin}(2007)}]{garaud07}
{Garaud}, P., \& {Lin}, D.~N.~C. 2007, \apj, 654, 606

\bibitem[{{Hartmann} {et~al.}(1998){Hartmann}, {Calvet}, {Gullbring}, \&
  {D'Alessio}}]{hartmann98}
{Hartmann}, L., {Calvet}, N., {Gullbring}, E., \& {D'Alessio}, P. 1998, \apj,
  495, 385

\bibitem[{{Hayashi}(1981)}]{hayashi81}
{Hayashi}, C. 1981, Progress of Theoretical Physics Supplement, 70, 35

\bibitem[{{Henning} \& {Semenov}(2013)}]{henning13}
{Henning}, T., \& {Semenov}, D. 2013, Chemical Reviews, 113, 9016

\bibitem[{{Hollenbach} {et~al.}(2009){Hollenbach}, {Kaufman}, {Bergin}, \&
  {Melnick}}]{hollenbach09}
{Hollenbach}, D., {Kaufman}, M.~J., {Bergin}, E.~A., \& {Melnick}, G.~J. 2009,
  \apj, 690, 1497

\bibitem[{{Hughes} \& {Armitage}(2010)}]{hughes10}
{Hughes}, A.~L.~H., \& {Armitage}, P.~J. 2010, \apj, 719, 1633

\bibitem[{{Ilgner} {et~al.}(2004){Ilgner}, {Henning}, {Markwick}, \&
  {Millar}}]{ilgner04}
{Ilgner}, M., {Henning}, T., {Markwick}, A.~J., \& {Millar}, T.~J. 2004, \aap,
  415, 643

\bibitem[{{Kennedy} {et~al.}(2006){Kennedy}, {Kenyon}, \&
  {Bromley}}]{kennedy06}
{Kennedy}, G.~M., {Kenyon}, S.~J., \& {Bromley}, B.~C. 2006, \apjl, 650, L139

\bibitem[{{Kraus} {et~al.}(2011){Kraus}, {Ireland}, {Martinache}, \&
  {Hillenbrand}}]{kraus11}
{Kraus}, A.~L., {Ireland}, M.~J., {Martinache}, F., \& {Hillenbrand}, L.~A.
  2011, \apj, 731, 8

\bibitem[{{Kreidberg} {et~al.}(2015){Kreidberg}, {Line}, {Bean}, {Stevenson},
  {Desert}, {Madhusudhan}, {Fortney}, {Barstow}, {Henry}, {Williamson}, \&
  {Showman}}]{kreidberg15}
{Kreidberg}, L., {Line}, M.~R., {Bean}, J.~L., {et~al.} 2015, ArXiv e-prints,
  arXiv:1504.05586

\bibitem[{{Krijt} {et~al.}(2015){Krijt}, {Ormel}, {Dominik}, \&
  {Tielens}}]{krijt15}
{Krijt}, S., {Ormel}, C.~W., {Dominik}, C., \& {Tielens}, A.~G.~G.~M. 2015,
  \aap, 574, A83

\bibitem[{{Lambrechts} \& {Johansen}(2012)}]{lambrechts12}
{Lambrechts}, M., \& {Johansen}, A. 2012, \aap, 544, A32

\bibitem[{{Madhusudhan} {et~al.}(2014){Madhusudhan}, {Amin}, \&
  {Kennedy}}]{madhu14}
{Madhusudhan}, N., {Amin}, M.~A., \& {Kennedy}, G.~M. 2014, \apjl, 794, L12

\bibitem[{{Madhusudhan} {et~al.}(2011){Madhusudhan}, {Harrington}, {Stevenson},
  {Nymeyer}, {Campo}, {Wheatley}, {Deming}, {Blecic}, {Hardy}, {Lust},
  {Anderson}, {Collier-Cameron}, {Britt}, {Bowman}, {Hebb}, {Hellier},
  {Maxted}, {Pollacco}, \& {West}}]{madhu11}
{Madhusudhan}, N., {Harrington}, J., {Stevenson}, K.~B., {et~al.} 2011, \nat,
  469, 64

\bibitem[{{Martin} \& {Livio}(2012)}]{martin12}
{Martin}, R.~G., \& {Livio}, M. 2012, \mnras, 425, L6

\bibitem[{{Mathews} {et~al.}(2013){Mathews}, {Klaassen}, {Juh{\'a}sz},
  {Harsono}, {Chapillon}, {van Dishoeck}, {Espada}, {de Gregorio-Monsalvo},
  {Hales}, {Hogerheijde}, {Mottram}, {Rawlings}, {Takahashi}, \&
  {Testi}}]{mathews13}
{Mathews}, G.~S., {Klaassen}, P.~D., {Juh{\'a}sz}, A., {et~al.} 2013, \aap,
  557, A132

\bibitem[{{Min} {et~al.}(2011){Min}, {Dullemond}, {Kama}, \& {Dominik}}]{min11}
{Min}, M., {Dullemond}, C.~P., {Kama}, M., \& {Dominik}, C. 2011, Icarus, 212,
  416

\bibitem[{{Mordasini} {et~al.}(2014){Mordasini}, {Klahr}, {Alibert}, {Miller},
  \& {Henning}}]{mordasini14}
{Mordasini}, C., {Klahr}, H., {Alibert}, Y., {Miller}, N., \& {Henning}, T.
  2014, \aap, 566, A141

\bibitem[{{Mulders} {et~al.}(2015){Mulders}, {Ciesla}, {Min}, \&
  {Pascucci}}]{mulders15}
{Mulders}, G.~D., {Ciesla}, F.~J., {Min}, M., \& {Pascucci}, I. 2015, \apj,
  807, 9

\bibitem[{{{\"O}berg} {et~al.}(2011){{\"O}berg}, {Murray-Clay}, \&
  {Bergin}}]{oberg11}
{{\"O}berg}, K.~I., {Murray-Clay}, R., \& {Bergin}, E.~A. 2011, \apjl, 743, L16

\bibitem[{{Okuzumi} {et~al.}(2012){Okuzumi}, {Tanaka}, {Kobayashi}, \&
  {Wada}}]{okuzumi12}
{Okuzumi}, S., {Tanaka}, H., {Kobayashi}, H., \& {Wada}, K. 2012, \apj, 752,
  106

\bibitem[{{P{\'e}rez} {et~al.}(2012){P{\'e}rez}, {Carpenter}, {Chandler},
  {Isella}, {Andrews}, {Ricci}, {Calvet}, {Corder}, {Deller}, {Dullemond},
  {Greaves}, {Harris}, {Henning}, {Kwon}, {Lazio}, {Linz}, {Mundy}, {Sargent},
  {Storm}, {Testi}, \& {Wilner}}]{perez12}
{P{\'e}rez}, L.~M., {Carpenter}, J.~M., {Chandler}, C.~J., {et~al.} 2012,
  \apjl, 760, L17

\bibitem[{{Piso} \& {Youdin}(2014)}]{piso14}
{Piso}, A.-M.~A., \& {Youdin}, A.~N. 2014, \apj, 786, 21

\bibitem[{{Podolak} \& {Zucker}(2004)}]{podolak04}
{Podolak}, M., \& {Zucker}, S. 2004, Meteoritics and Planetary Science, 39,
  1859

\bibitem[{{Pollack} {et~al.}(1996){Pollack}, {Hubickyj}, {Bodenheimer},
  {Lissauer}, {Podolak}, \& {Greenzweig}}]{pollack96}
{Pollack}, J.~B., {Hubickyj}, O., {Bodenheimer}, P., {et~al.} 1996, Icarus,
  124, 62

\bibitem[{{Qi} {et~al.}(2013){Qi}, {{\"O}berg}, {Wilner}, {D'Alessio},
  {Bergin}, {Andrews}, {Blake}, {Hogerheijde}, \& {van Dishoeck}}]{qi13}
{Qi}, C., {{\"O}berg}, K.~I., {Wilner}, D.~J., {et~al.} 2013, Science, 341, 630

\bibitem[{{Rafikov}(2006)}]{rafikov06}
{Rafikov}, R.~R. 2006, \apj, 648, 666

\bibitem[{{Ricci} {et~al.}(2010){Ricci}, {Testi}, {Natta}, \&
  {Brooks}}]{ricci10}
{Ricci}, L., {Testi}, L., {Natta}, A., \& {Brooks}, K.~J. 2010, \aap, 521, A66

\bibitem[{{Shakura} \& {Sunyaev}(1973)}]{shakura73}
{Shakura}, N.~I., \& {Sunyaev}, R.~A. 1973, \aap, 24, 337

\bibitem[{{Sicilia-Aguilar} {et~al.}(2010){Sicilia-Aguilar}, {Henning}, \&
  {Hartmann}}]{sicilia10}
{Sicilia-Aguilar}, A., {Henning}, T., \& {Hartmann}, L.~W. 2010, \apj, 710, 597

\bibitem[{{Stevenson} {et~al.}(2014){Stevenson}, {Bean}, {Seifahrt},
  {D{\'e}sert}, {Madhusudhan}, {Bergmann}, {Kreidberg}, \&
  {Homeier}}]{stevenson14}
{Stevenson}, K.~B., {Bean}, J.~L., {Seifahrt}, A., {et~al.} 2014, \aj, 147, 161

\bibitem[{{Thiabaud} {et~al.}(2015){Thiabaud}, {Marboeuf}, {Alibert}, {Leya},
  \& {Mezger}}]{thiabaud15}
{Thiabaud}, A., {Marboeuf}, U., {Alibert}, Y., {Leya}, I., \& {Mezger}, K.
  2015, \aap, 574, A138

\bibitem[{{van der Marel} {et~al.}(2015){van der Marel}, {van Dishoeck},
  {Bruderer}, {P{\'e}rez}, \& {Isella}}]{vandermarel15}
{van der Marel}, N., {van Dishoeck}, E.~F., {Bruderer}, S., {P{\'e}rez}, L., \&
  {Isella}, A. 2015, \aap, 579, A106

\bibitem[{{van Dishoeck}(2006)}]{vandishoeck06}
{van Dishoeck}, E.~F. 2006, Proceedings of the National Academy of Science,
  103, 12249

\bibitem[{{Weidenschilling}(1977)}]{weidenschilling77}
{Weidenschilling}, S.~J. 1977, \mnras, 180, 57

\bibitem[{{Wilner} {et~al.}(2000){Wilner}, {Ho}, {Kastner}, \&
  {Rodr{\'{\i}}guez}}]{wilner00}
{Wilner}, D.~J., {Ho}, P.~T.~P., {Kastner}, J.~H., \& {Rodr{\'{\i}}guez}, L.~F.
  2000, \apjl, 534, L101

\bibitem[{{Youdin} \& {Lithwick}(2007)}]{youdin07}
{Youdin}, A.~N., \& {Lithwick}, Y. 2007, Icarus, 192, 588

\bibitem[{{Zhang} {et~al.}(2013){Zhang}, {Pontoppidan}, {Salyk}, \&
  {Blake}}]{zhang13}
{Zhang}, K., {Pontoppidan}, K.~M., {Salyk}, C., \& {Blake}, G.~A. 2013, \apj,
  766, 82

\bibitem[{{Zsom} {et~al.}(2010){Zsom}, {Ormel}, {G{\"u}ttler}, {Blum}, \&
  {Dullemond}}]{zsom10}
{Zsom}, A., {Ormel}, C.~W., {G{\"u}ttler}, C., {Blum}, J., \& {Dullemond},
  C.~P. 2010, \aap, 513, A57

\end{thebibliography}


\begin{thebibliography}
\end{thebibliography}
\fi

\if\bibinc y

\fi

\end{document}